\documentclass[a4paper, 11pt, oneside]{article} 

\usepackage{geometry}
\usepackage{booktabs} 

\usepackage{mathptmx} 

\usepackage{fancyhdr}
\usepackage[normalem]{ulem}
\usepackage{microtype}
\usepackage{graphicx}
\usepackage{colortbl}
\definecolor{mygray}{gray}{.88}
\usepackage{multirow}
\usepackage{amsmath}
\usepackage{bm}
\usepackage{epsfig}
\usepackage{authblk}
\usepackage{cite}
\usepackage{amsmath,amssymb,amsfonts}
\usepackage{algorithmic}
\usepackage{graphicx}
\usepackage{textcomp}
\usepackage{xcolor}
\usepackage{threeparttablex}
\usepackage{subcaption}
\usepackage{verbatim}
\usepackage{float}

\usepackage{array}
\renewcommand{\arraystretch}{1.2}
\newcolumntype{?}{!{\vrule width 0.7pt}}

\usepackage{multirow}
\usepackage{multicol}

\usepackage[font=small,labelfont=bf,tableposition=top]{caption}
\DeclareCaptionLabelFormat{andtable}{#1~#2  \&  \tablename~\thetable}

\def\BibTeX{{\rm B\kern-.05em{\sc i\kern-.025em b}\kern-.08em
    T\kern-.1667em\lower.7ex\hbox{E}\kern-.125emX}}

\usepackage[hyphens]{url}

\usepackage{color}
\usepackage{soul}
\usepackage{bbding}
\definecolor{mygray}{gray}{.88}

\begin{document} 

\begin{titlepage} 

	\centering 
	
	\scshape 
	
	\vspace*{\baselineskip} 
	
	
	\rule{\textwidth}{1.6pt}\vspace*{-\baselineskip}\vspace*{2pt} 
	\rule{\textwidth}{0.4pt} 
	
	\vspace{0.75\baselineskip} 
	
	{\LARGE HPC AI500:\\ The Methodology, Tools,  Roofline Performance Models, and Metrics for Benchmarking HPC AI Systems} 
	
	\vspace{0.75\baselineskip} 

	\rule{\textwidth}{0.4pt}\vspace*{-\baselineskip}\vspace{3.2pt} 
	\rule{\textwidth}{1.6pt} 
	
	\vspace{2\baselineskip} 
	
	
	
	\vspace*{3\baselineskip} 
	
	
	AUTHORS' CONTRIBUTION
	
	\vspace{0.5\baselineskip} 
	
	{\scshape SECTION 1 IS CONTRIBUTED BY JIANFENG ZHAN AND ZIHAN JIANG. SECTION 2 IS CONTRIBUTED BY JIANFENG ZHAN, ZIHAN JIANG, AND FEI TANG. SECTION 3 IS CONTRIBUTED BY JIANFENG ZHAN. SECTION 4 IS CONTRIBUTED BY XINGWANG XIONG, ZIHAN JIANG, LEI WANG, WANLING GAO, AND JIANFENG ZHAN. SECTION 5 IS CONTRIBUTED BY ZIHAN JIANG, LEI WANG, CHUNJIE LUO, WANLING GAO, JIANFENG ZHAN, AND HONGXIAO LI. SECTION 6 IS CONTRIBUTED BY LEI WANG, ZIHAN JIANG, WANLING GAO, AND JIANFENG ZHAN. SECTION 7 IS CONTRIBUTED BY ZIHAN JIANG, XINGWANG XIONG, LEI WANG, WANLING GAO, CHUNXIN LAN, AND JIANFENG ZHAN. SECTION 8 IS CONTRIBUTED BY ZIHAN JIANG, LEI WANG, WANLING GAO, AND JIANFENG ZHAN. SECTION 9 ISCONTRIBUTED BY JIANFENG ZHAN.\\ }
	
	
	\vspace{0.5\baselineskip} 

	\vfill 
	
	
	\epsfig{file=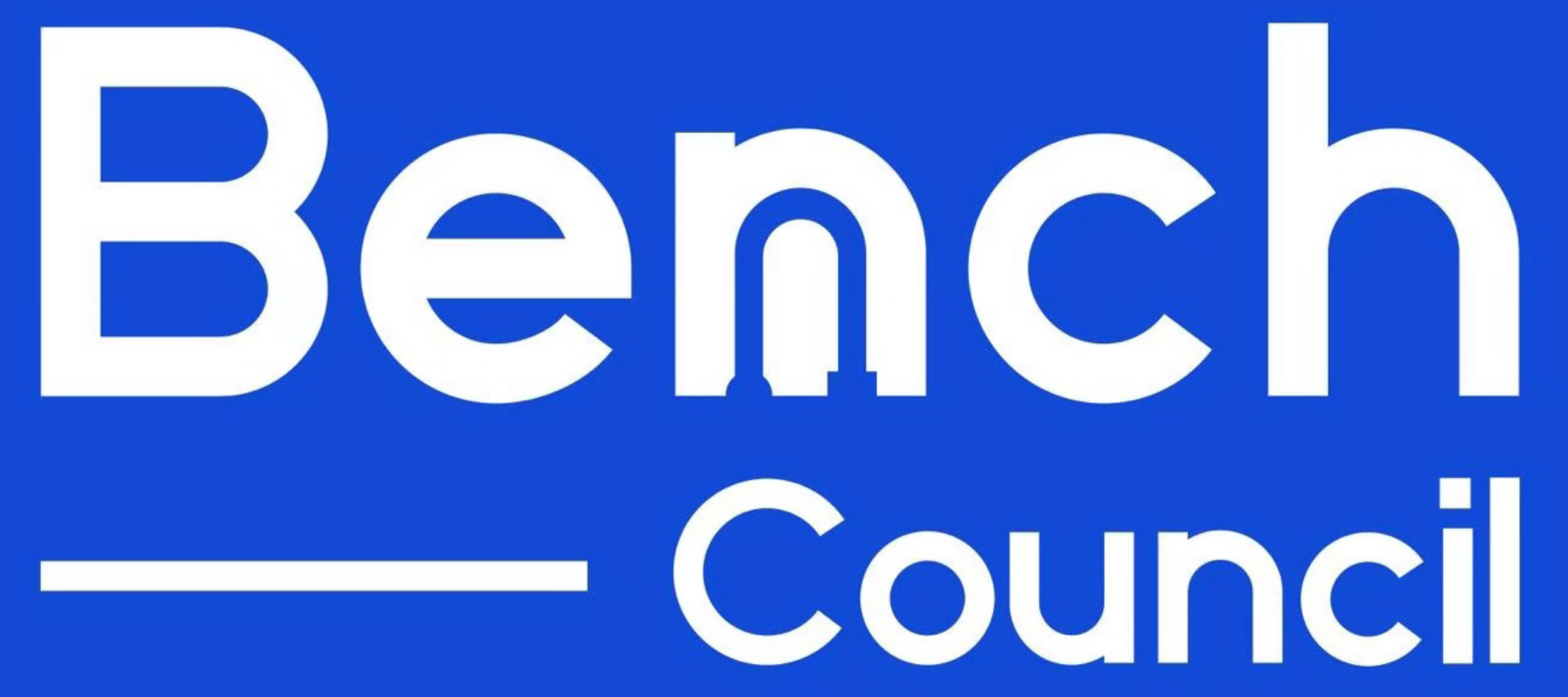,height=2cm}
	\textit{\\BenchCouncil: International Open Benchmarking Council\\Chinese Academy of Sciences\\Beijing, China\\http://www.benchcouncil.org/HPCAI500/index.html} 
	\vspace{5\baselineskip} 

	Technical Report No. BenchCouncil-HPCAI500-2020-1 
	
	{\large June 30, 2020} 

\end{titlepage}


\title{HPC AI500: The Methodology, Tools, Roofline Performance Models, and Metrics for Benchmarking HPC AI Systems*}

\author[1,3]{Zihan Jiang}
\author[1,2]{Lei Wang}
\author[1,3]{Xingwang Xiong}
\author[1,2]{Wanling Gao}
\author[1,2]{Chunjie Luo}
\author[1,3]{Fei Tang}
\author[1]{Chuanxin Lan}
\author[1,3]{Hongxiao Li}
\author[1,2,3]{Jianfeng Zhan\thanks{Jianfeng Zhan is the corresponding author.}}

\affil[1]{State Key Laboratory of Computer Architecture, Institute of Computing Technology, Chinese Academy of Sciences \\ \{jiangzihan, wanglei\_2011, xingwangxiong, gaowanling, luochunjie, lanchuanxin, tangfei, lihongxiao, zhanjianfeng\}@ict.ac.cn}
\affil[2]{BenchCouncil (International Open Benchmarking Council)}
\affil[3]{University of Chinese Academy of Sciences}

\date{June 30, 2020}
\maketitle

\section{Abstract}

The recent years witness a trend of applying large-scale distributed deep learning algorithms in both business and scientific computing areas, whose goal is to speed up the training time to achieve a state-of-the-art quality. The HPC community feels a great interest in building the HPC AI systems that are dedicated to running those workloads. The HPC AI benchmarks accelerate the process. Unfortunately, benchmarking HPC AI systems at scale raises serious challenges. None of previous HPC AI benchmarks achieve the goal of being equivalent, relevant, representative, affordable, and repeatable.

This paper presents a comprehensive methodology, tools, Roofline performance models, and innovative metrics for benchmarking, optimizing, and ranking HPC AI systems, which we call HPC AI500 V2.0. We abstract the HPC AI system into nine independent layers, and present  explicit benchmarking rules and procedures to assure  equivalence of each layer, repeatability, and replicability. On the basis of AIBench--by far the most comprehensive AI benchmarks suite, we present and build two HPC AI benchmarks from both business and scientific computing: Image Classification, and  Extreme Weather Analytics, achieving both representativeness and affordability. To ranking the performance and energy-efficiency of HPC AI systems, we propose Valid FLOPS, and Valid FLOPS per watt, which impose a penalty on failing to achieve the target quality. We propose using convolution and GEMM--- the two most intensively-used kernel functions of AIBench to measure the upper bound performance  of the HPC AI systems, and present HPC AI roofline models for guiding performance optimizations. The evaluations show our methodology, benchmarks, performance models, and metrics can measure, optimize, and rank  the HPC AI systems in a scalable, simple, and affordable way. The specification, source code, datasets, and benchmarking data are publicly available from \url{http://www.benchcouncil.org/benchhub/hpc-ai500-benchmark} .

\section{Introduction}

\begin{figure}[ht]
  \centering
  \includegraphics[width=\linewidth]{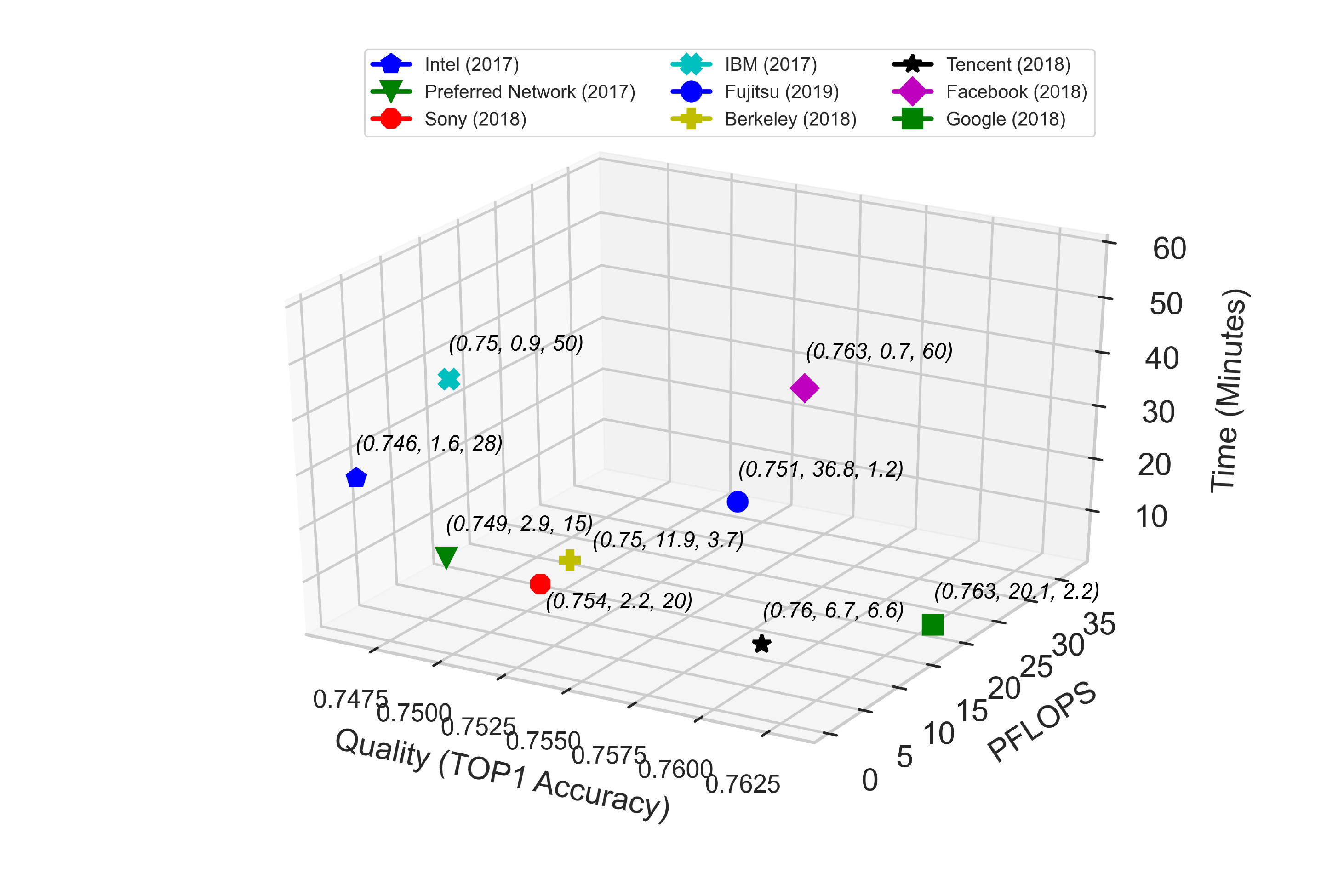}
  \caption{ImageNet/ResNet-50 training is one well-known showcase for optimizing HPC AI systems. It reports the performance in terms of a ternary tuple (achieved quality, PFLOPS, time-to-quality--minutes). The past witnesses the systems performance varying wildly from (74.6\%, 1.6, 28) to (75.1\%, 36.8, 1.2). Table~\ref{table:imagenet-and-hpcai} summarizes the  utilized  optimization  approaches. As no equivalent benchmarking rule is stated, we can not objectively derive the  performance edge of one system against the others. }

  \label{fig:hpcai-trend}
\end{figure}

The huge success of AlexNet~\cite{krizhevsky2012imagenet} in the ImageNet~\cite{deng2009imagenet} competition marks the booming success of  deep learning (DL) in a wide range of commercial application areas.
Many commercial fields, like image recognition, and natural language processing achieve unprecedented accuracy, even outperforming common human being's capability. Though it is much challenging to obtain high quality labeled scientific data sets, there is an increasing trend in applying DL in scientific computing areas~\cite{dlinscience,ravanbakhsh2016estimating,liu2016application,mathuriya2018cosmoflow}.


With massive training data available, the recent years witness a trend of applying distributed DL algorithms at scale in both commercial and scientific computing areas. Motivated by these emerging HPC AI workloads, the HPC community feels a great interest in building HPC AI systems to reduce time-to-quality--the training time to achieve a convergent quality. For example, the Summit system~\cite{kurth2018exascale} is built to tackle huge AI challenges. 
The benchmark accelerates the process~\cite{hennessy2011computer,mattson2019mlperf}, as it provides not only design inputs, but also evaluation and optimization metric and methodology~\cite{tang2020aibench, gao2020aibench}. However, there are several challenges in benchmarking HPC AI systems. 

First, it is nontrivial to prove the equivalence of two AI benchmark implementations on different systems or even the same system with different scales. Equivalence quantifies how equivalent two benchmarks implementations on different systems or the same system with different scales.  There are complex interactions among hardware and  software systems, which is further aggravated by the AI algorithm complexity. Even for the same AI algorithms, there are huge parameters significantly impacting learning dynamics~\cite{mattson2019mlperf}.  ImageNet/ResNet-50 (Image Classification) training is one well-known showcase for optimizing HPC AI systems. Table~\ref{table:imagenet-and-hpcai} summarizes the state-of-the-art and state-of-the-practice optimization approaches in ImageNet/ResNet-50 training. Unfortunately,  without equivalent benchmarking rules explicitly stated, 
we can not objectively derive the  performance edge of one system against the others from Fig.~\ref{fig:hpcai-trend}.

The second challenge inherits from the the conflict of two classical benchmarking methodologies with the emphasis of different requirements. 
On one hand, as no single benchmark or metric can measure the performance of computer systems on all applications~\cite{gray1993database}, being relevant, representative, and diverse is of paramount importance~\cite{tang2020aibench}. 
On the other hand,  TOP500~\cite{dongarra1997top500} establishes the de facto super computer benchmark standard in terms of three defining characteristics: scalable, simple, and affordable.

\begin{figure}[!ht]

  \centering
  \includegraphics[width=0.8\linewidth]{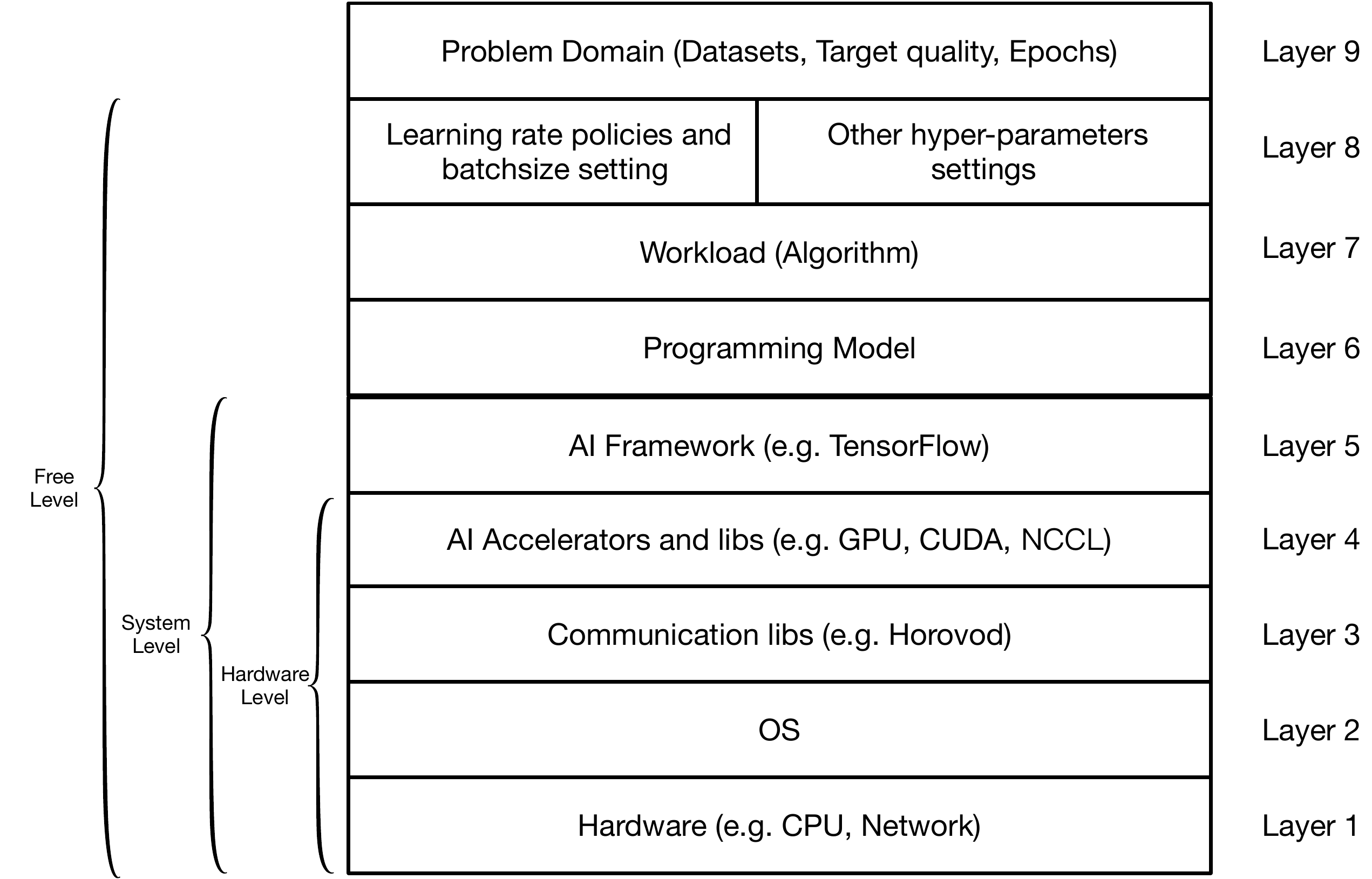}
  \caption{The equivalent perspective of HPC AI500 V2.0 Methodology. We abstract the HPC AI system into nine independent layers: put each layer under test while keeping other layers intact. We provide three high levels of benchmarking: hardware, system, and free:  put the related layers together under test while keeping other layers intact with only allowed changes stated in the benchmarking rules.}
  \label{fig:nine-layer}
\end{figure}

In the AI domain, there are massive AI tasks and  models with different performance metrics. For example, by far the most comprehensive and representative AI benchmark suite--AIBench~\cite{tang2020aibench, gao2018aibench, gao2019aibench, gao2020aibench} contains seventeen AI tasks.  It is not affordable to implement so many massive benchmarks and further perform benchmarking at scale. So what are the criteria for deciding the benchmarks that can fairly and objectively measure the HPC AI systems. 

Third, the benchmark mandates being repeatable, while the nature of AI is stochastic, allowing
multiple different but equally valid solutions~\cite{mattson2019mlperf}. The uncertainty of HPC AI is manifested by run-to-run variation in terms of epochs-to-quality and the effect of scaling training on time-to-quality~\cite{mattson2019mlperf,zhan2019benchcouncil}. For the first time, Tang et al.~\cite{tang2020aibench} quantify the variations of seventeen AI benchmarks of AIBench. 
They found that the run-to-run variations vary from 0\% to 38.46\%  in terms of the ratio of the standard deviation to the mean of the training epochs to achieve a convergent quality.  

None of previous HPC AI benchmarks achieve the goal of being equivalent, relevant, representative, affordable, and repeatable.
They either are not representative or even irrelevant to HPC AI workloads in terms of kernel functions~\cite{hplai,micikevicius2017mixed}, or overlook the differences of HPC AI workloads between scientific and business computing~\cite{mattson2019mlperf}, or fail to specify fair and equivalent benchmarking rules across different HPC AI systems~\cite{mattson2019mlperf}. Moreover, they fail to propose simple and AI domain-specific metric to score and rank HPC AI systems.

The  micro  benchmark  like HPL-AI~\cite{micikevicius2017mixed}, which only contains LU decomposition, is affordable to perform a fair comparison of  competing  systems  by  isolating  hardware  and  software from statistical optimizations~\cite{mattson2019mlperf}. However, we found it is irrelevant to most of AI workloads in Section~\ref{subsec:challenge_relevance}. Moreover, the traditional micro or kernel benchmarking methodology,  widely used in the HPC communities, can lead to misleading conclusion, as the mixed precision optimizations indeed improve the FLOPS of a micro benchmark like convolution,  while significantly impact  time-to-quality 
of an AI task like image classification as discussed in Section~\ref{subsec:challenge_relevance}. 

This paper presents HPC AI500 V2.0--a comprhensieve HPC AI benchmarking methodology, tools, performance models, and metrics.
As shown in Fig.~\ref{fig:nine-layer}, we abstract the HPC AI system into nine independent layers. 
To perform fair benchmarking across different systems or the same system with different scales, we present explicit benchmarking rules to assure  equivalence of each layer, repeatability, and replicability of those two benchmarks. We put each layer under test while keeping the other layers intact. Also, We propose three high levels of benchmarking: hardware, system, and free (Fig.~\ref{fig:nine-layer}): put the related layers under test while keeping the other layers intact unless otherwise stated.

On the basis of AIBench
, we present two 
benchmarks: Image Classification with state-of-the-art quality on the ImageNet dataset (business computing), and Extreme Weather Analytics (EWA) with state-of-the-art quality on the  EWA dataset (scientific computing) to measure HPC AI systems. These two benchmarks represent two  clusters of AI benchmarks--thirteen AI benchmarks from AIBench from perspectives of computing areas (business vs. scientific computing), diversity of model complexity (from 0.03 million to 68.39 million in terms of model parameters ), computational cost (from 0.09 MFLOPs to 157.80 GFLOPs in terms of a single forward computation), and convergence rate (from 6 epochs to 304 epochs). Moreover, our decision also takes into account their repeatablility, and whether these benchmarks have widely-accepted metrics or not.

 To rank HPC AI systems, 
we propose two metrics, named Valid FLOPS, and Valid FLOPS per watt, to emphasise the vital importance of achieving the state-of-the-art quality, and an auxiliary metrics--time-to-quality.

 We propose using convolution and GEMM (GEneral Matrix to Matrix Multiplication)--two most intensively-used kernel functions of AIBench to measure the upper bound performance  of the HPC AI systems, and present corresponding single-node and distributed-system HPC AI roofline models for guiding performance optimizations.

The evaluations show our benchmarks can fairly measure the HPC AI systems in a scalable, simple, and affordable way. Our Roofline models are helpful to system optimizations. Our metrics can be used to rank HPC AI systems in a simple and visual manner.

\section{The challenges of HPC AI benchmarking}~\label{challenge}

The challenges of HPC AI benchmarking inherit from the complexity of benchmarking scalable hardware and software systems, which are further exaggerated by the uncertainty of AI algorithms.

\subsection{Equivalence}\label{subsec:challenge-equivalence}

\begin{table}[!ht]
\scriptsize
\caption{The summary of  the  utilized  optimization  approaches in ImageNet/ResNet-50 training. The optimization approaches of each system are inconsistent or inequivalent. Please note that only the optimizations items ~\emph{in italics} are allowed to change in the HPC AI500 benchmarking rules (defined in Section~\ref{sec:benchmarking-rules}).}
    \centering
    \begin{tabular}{m{0.07\textwidth}?>{\centering}m{0.08\textwidth}|>{\centering}m{0.13\textwidth}|>{\centering}m{0.08\textwidth}|>{\centering}m{0.08\textwidth}?>{\centering}m{0.1\textwidth}|>{\centering}m{0.11\textwidth}|>{\centering}m{0.1\textwidth}|>{\centering\arraybackslash}m{0.1\textwidth}}
    \toprule
        & \multicolumn{4}{c?}{\textbf{System-level}} & \multicolumn{4}{c}{\textbf{Algorithm-level}} \\
        \cmidrule(lr){2-5} \cmidrule(lr){6-9}
        & \emph{parallel Mode} & \emph{Communication} & \emph{Precision} & \emph{Data Staging} & \emph{Learning Rate Policy} & Data Argumentation & Model Architecture& Others \\
    \midrule
    \textbf{Facebook}\cite{goyal2017accurate} & Data parallelism & Recursive halving and doubling and ring all-reduce. & N/A  & N/A  & Linear scaling and warm up~\cite{krizhevsky2014one}. & N/A & N/A  & Momentum correction; Data shuffling based on the workers.\\\midrule
    \textbf{Intel}\cite{codreanu2017scale} & Data parallelism & Intel MLSL~\cite{sridharan2018scale} & N/A  & N/A  & Linear scaling and warmup; final collapse. & N/A & N/A  & Collapsed ensembles; Dynamically change weight decay.\\\midrule
    \textbf{IBM}\cite{cho2017powerai}  & Data parallelism & Topology aware & N/A  & N/A   & Linear scaling, warmup~\cite{krizhevsky2014one} & N/A & N/A  & Momentum correction; Data shuffling based on the workers. \\\midrule
    \textbf{Berkeley}\cite{10.1145/3225058.3225069}  & Data parallelism & Intel MLSL~\cite{intel} & N/A  & N/A & Linear scaling and warmup~\cite{krizhevsky2014one}; LARS~\cite{you2017imagenet}. & N/A & N/A  & N/A  \\\midrule
    \textbf{Preferred Networks}\cite{akiba2017extremely}  & Data parallelism & Ring all-reduce; Communication compression. & N/A  & N/A   & Linear scaling, RMSprop warmup, and slow-start; & N/A & Batch normalization: without moving averages. & N/A  \\\midrule
    \textbf{Sony}\cite{tanakaimagenet}  & Data parallelism & 2D-Torus all-reduce; Communication compression; Communication tensor fusion. & Mixed precision training: FP16 \& FP32  & N/A & Linear scaling and warmup~\cite{krizhevsky2014one}; LARS~\cite{you2017imagenet} & Adding, scaling, rotations ,etc & Batch normalization: without moving averages. & N/A \\\midrule
    \textbf{Tencent}\cite{jia2018highly}  & Data parallelism & Hierarchical all-reduce; Communication compression; Communication tensor fusion. & Mixed precision training: FP16 \& FP32  & Efficient input pipeline & Linear scaling and warmup~\cite{krizhevsky2014one}; LARS~\cite{you2017imagenet} & N/A & Batch normalization: eliminating weight decay. & N/A  \\\midrule
    \textbf{Google}\cite{ying2018image}  & Data parallelism & 2D-Mesh all-reduce; & Mixed precision training: BFLOAT16~\cite{googlebfloat} \& FP32.  & Efficient input pipeline &  Linear scaling and warmup~\cite{krizhevsky2014one}; LARS~\cite{you2017imagenet} &  Fused JPEG decode and cropping & Distributed batch normalization & N/A  \\\midrule
    \textbf{Fujitsu}\cite{yamazaki2019yet}  & Data parallelism & Communication tensor fusion; Optimal scheduling by grouping layers; Calculate the norms of layers in parallel.& Mixed precision training: FP16 \& FP32.  & N/A  & Linear scaling and warmup~\cite{krizhevsky2014one}; LARS~\cite{you2017imagenet}  & N/A & N/A  & Label smoothing~\cite{szegedy2016rethinking} \\
    \bottomrule
    \end{tabular}
    \label{table:imagenet-and-hpcai}
\end{table}

For the same AI algorithms, there are huge parameters significantly impacting learning dynamics~\cite{mattson2019mlperf}. Even for the same system with different scales, the interactions among system size, minibatch size, and learning dynamics have a significant impact on time-to-quality and computation overhead in terms of FLOPS~\cite{mattson2019mlperf,goyal2017accurate,you2017imagenet}. So for the same AI task, it is non-trivial to prove the equivalence of two benchmark implementations on different systems or even the same system with different scales. 

ImageNet/ResNet-50 training is one widely-used showcase for optimizaing HPC AI systems. Fig.~\ref{fig:hpcai-trend} shows the systems performance varies wildly: the performance gap in terms of FLOPS is 50x. Accordingly,  Table~\ref{table:imagenet-and-hpcai} summarizes the state-of-the-art and state-of-the-practice work on ImageNet training at scale. 
In addition to the system-level optimizations (e.g. more efficient communication typologies), some algorithm-level optimizations involve changing model architectures (e.g. optimizations on batch normalization) or learning rate policies, i.e., LARS~\cite{you2017imagenet}. As there are prohibitively complex interactions among  hardware systems, software systems, and algorithms, previous work fails to clearly state the equivalent rules of each hardware or software layer for benchmarking HPC AI systems.

\subsection{Representative,  Affordable, and Relevant}\label{subsec:challenge_relevance}

The second challenge inherits from the the conflict of two classical benchmarking methodologies with the emphasis of different requirements. 

On one hand, the SPECCPU~\cite{speccpu}, PARSEC~\cite{bienia2008parsec}, and TPC benchmarks, like TPC-DS~\cite{tpcds} witness the paramount importance~\cite{tang2020aibench} of being representative and diverse,  as no single benchmark or metric can measure the performance of computer systems on all applications~\cite{gray1993database}. 

On the other hand, TOP500~\cite{dongarra1997top500} defines three distinctive characteristics of  the de facto super computer benchmark standard: affordable, simple, and scalable.  Affordable has two implications: first, the benchmark is easy to port to a new system or architecture; second, the benchmarking cost is affordable for measuring a systems at scale. Simple indicates the number of the metric is not only linear, orthogonal, and monotony~\cite{dongarra1997top500}, but also easily interpretable and understandable. Scalable means  the benchmark can be used to measured different scales of system,
and the problem size can be scaled up and down.



In the AI domain, there are massive AI tasks and models with different performance metrics. For example, AIBench~\cite{tang2020aibench}
contains seventeen representative AI tasks, including Image Classification, Object Detection, Learning to Rank, Image Generation, Text-to-Text Translation, Image-to-Text, Image-to-Image Translation, Speech Recognition, Face Embedding, 3D Face Recognition, Recommendation, Video Prediction, Image Compression, 3D Object Reconstruction, Text Summarization, Spatial Transformer, and Neural Architecture Search. For HPC AI benchmarking,  it is not affordable to implement so many massive benchmarks and further perform benchmarking at scale. 

\begin{figure}[!ht]
\begin{subfigure}{.48\textwidth}
  \centering
  \includegraphics[width=.6\linewidth,height=0.5\linewidth]{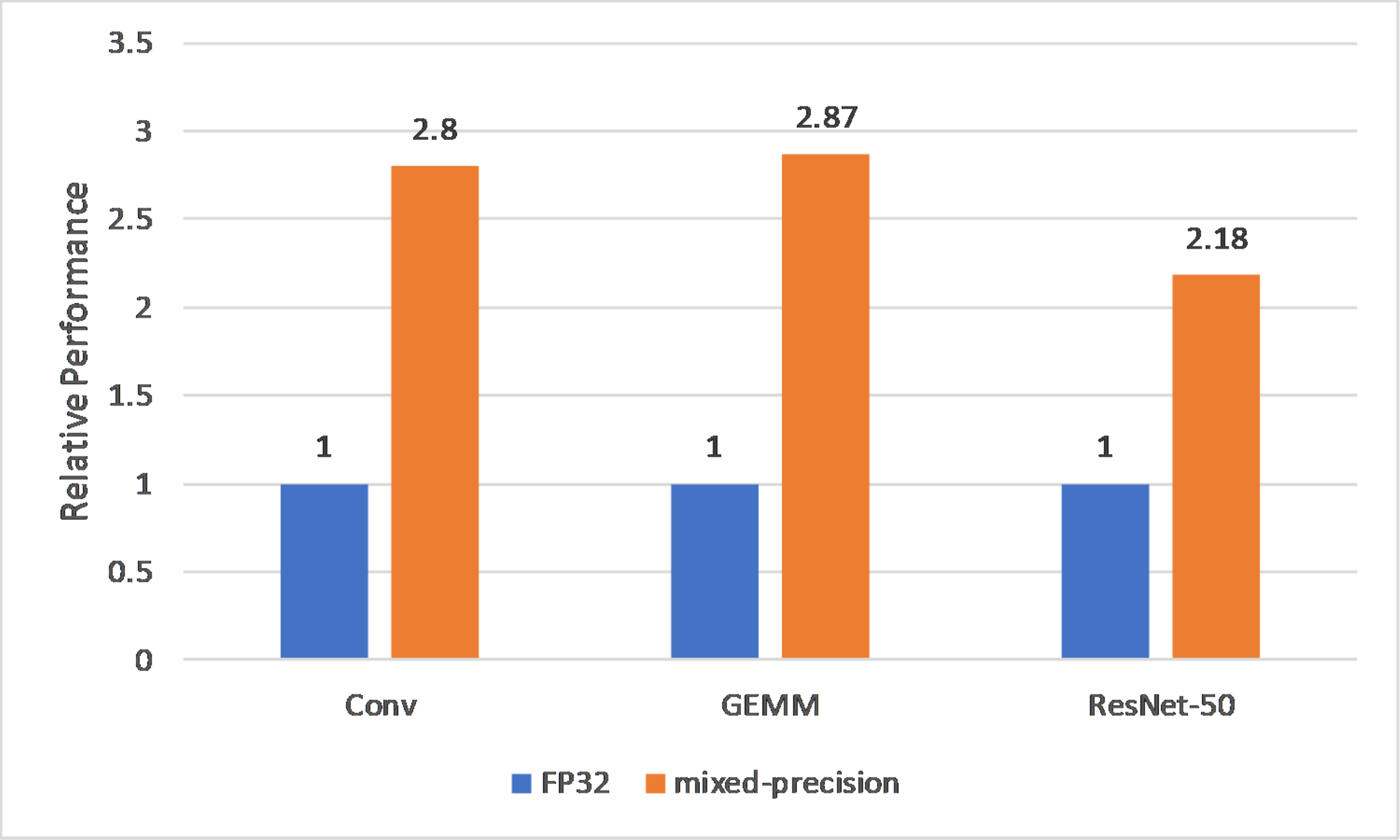}  
  \label{fig:mixed-precision-improvement}
\end{subfigure}
\begin{subfigure}{.48\textwidth}
  \centering
  \includegraphics[width=.8\linewidth,height=0.55\linewidth]{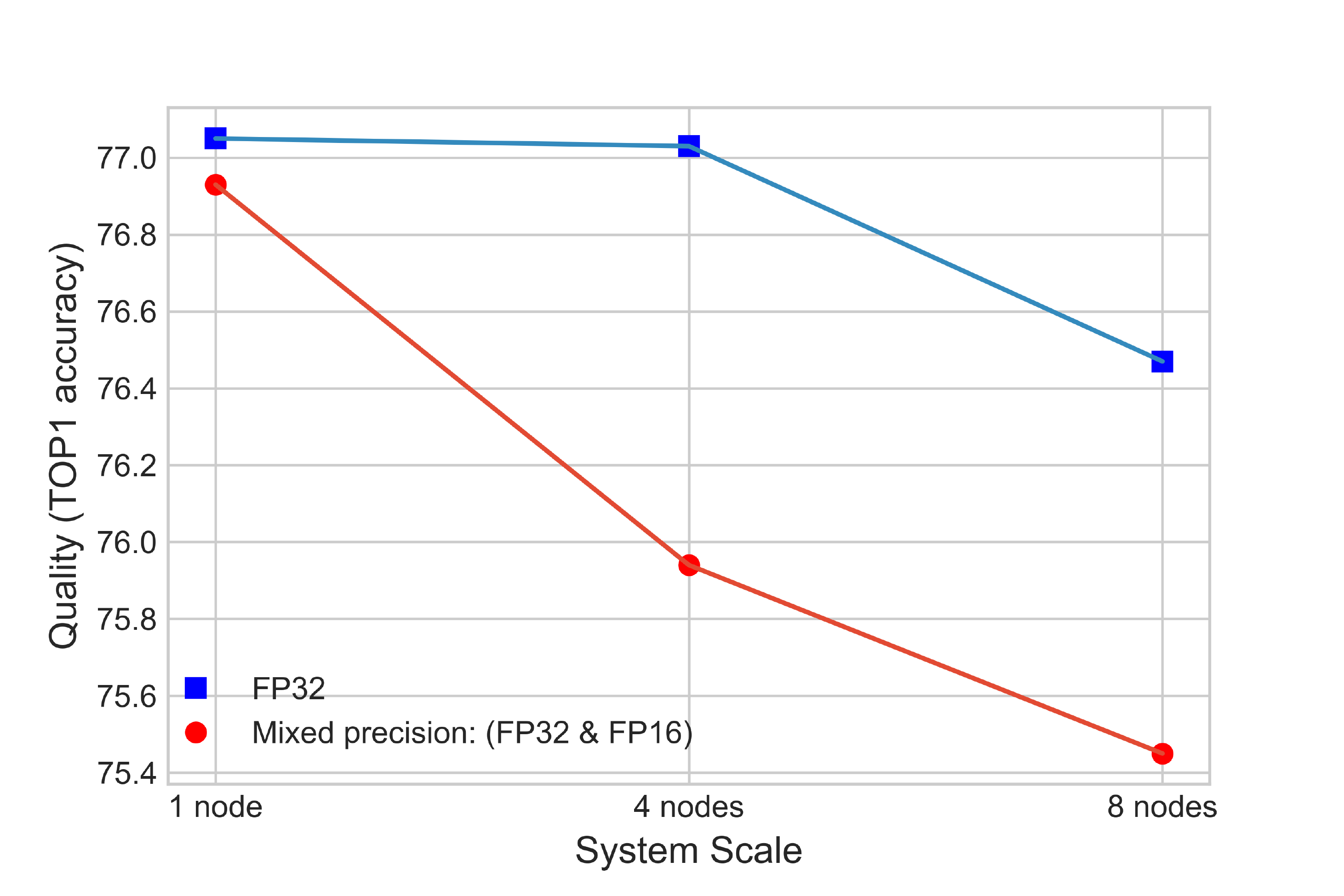}  
  \label{fig:mixed-accuracy-drop}
\end{subfigure}
\caption{With respect to the FP32 implementation, the mixed precision one   speeds up 2x the FLOPS of two micro benchmarks: Conv and GEMM and a component benchmark: ResNet-50 (LEFT), while incurring deteriorated accuracy drop of ResNet-50  when the system scale increases  (Right):  0.12\% at 1 node while  about 1\% at 8 nodes.}

\label{fig:micro_mixed_optimization}
\end{figure}

The traditional micro or kernel benchmarking methodology, which is  widely in the HPC communities, can lead to misleading conclusion, as the mixed precision optimizations indeed improve the FLOPS of a micro benchmark like convolution, while significantly impact time-to-quality
of an AI task like Image Classification. Fig.~\ref{fig:micro_mixed_optimization} shows that the mixed precision implementation increases the FLOPS of both micro and component benchmarks, while incurring accuracy drop as the system scale increases.

\begin{figure}[!ht]
  \centering
  \includegraphics[width=.6\linewidth]{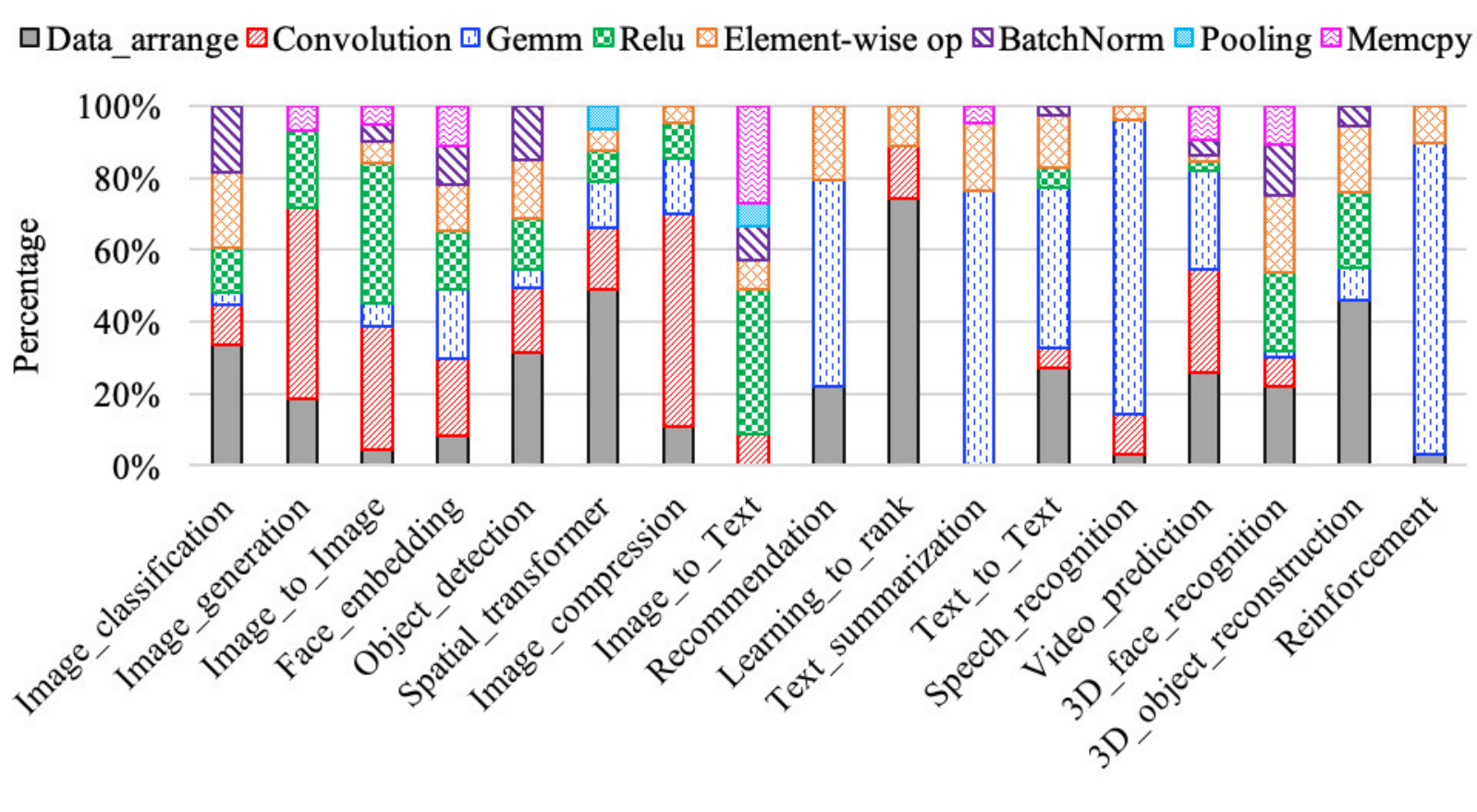}
  \caption{The kernel function breakdown of the 17 representative AI workloads from AIBench~\cite{tang2020aibench}, indicating the LU factorization is irrelevant. }
  \label{fig:breakdown_AIBench}
\end{figure}

Last but not least, the relevancy~\cite{gray1993benchmark} of a benchmark indicates that it must measure the peak performance and price/performance of systems when performing typical operations within that problem domain.
The  micro benchmark like HPL-AI~\cite{micikevicius2017mixed}, which only contains LU decomposition, is affordable to perform a fair comparison of competing  systems  by isolating hardware and software from statistical  optimizations~\cite{mattson2019mlperf}. However, we found it is irrelevant to most of AI workloads in AIBench. As shown in Fig. ~\ref{fig:breakdown_AIBench}, the dominated kernel functions are convolution and matrix multiplication.

\subsection{Repeatability}\label{subsec:challenge_repeatablity} 

  Repeatability~\cite{RRRultrasound, ACM_defintion} refers to the variation in repeat measurements of different runs of the same benchmark implementation,  by the same team, on the same system under  the identical configurations. 

\begin{figure}[!ht]
\centering
\includegraphics[width=.6\linewidth]{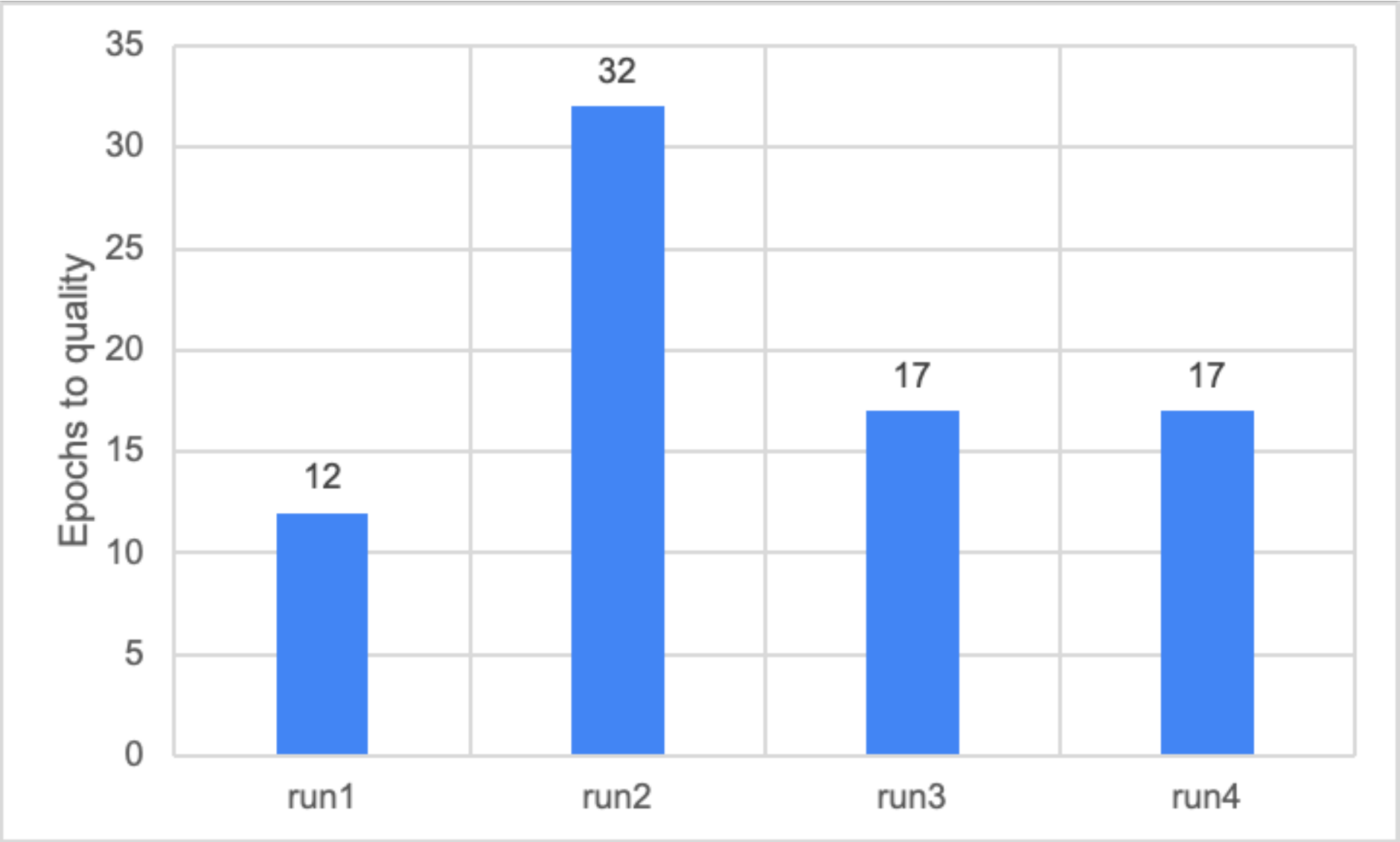}
\caption{The worst unrepeatable benchmark from AIBench is 3D Face Recognition. Its run-to-run variation is high as 38.46\%. The variation is defined as the ratio of the standard deviation to the mean of the training epochs to the achieved quality~\cite{tang2020aibench}.} 
\label{fig:run_to_run_variance}
\end{figure}

 Table~\ref{fig:AIBench-randomness} shows run-to-run variations of 17 benchmarks from AIBench varying from 0\% to 38.46\%. As shown in Fig.~\ref{fig:run_to_run_variance},
the variation of 
 3d Face Recognition is high as 38.46\%. There are diverse reasons for the uncertainty of different benchmarks. For NAS (network architecture searching), it constructs the network architecture by randomly sampling building blocks (e.g. convolution) from a predefined search space. 
 In addition, the complex design itself, which involves many hyper-parameters, makes AutoML hard to evaluate~\cite{yang2019evaluation}.

\begin{table}[ht]
\scriptsize 
\caption{The run-to-run variations of seventeen AI benchmarks of  AIBench~\cite{tang2020aibench}}
\renewcommand\arraystretch{1.2}
\label{fig:AIBench-randomness}
\center 
\begin{tabular}{|p{1in}|p{1.5in}|p{1in}|p{1in}|}
\hline
\textbf{No.} & \textbf{Component Benchmark} & \textbf{Variation} & \textbf{Repeat Times} \\
\hline
DC-AI-C1 & Image Classification & 1.12\% & 5 \\
\hline
DC-AI-C2 & Image Generation & Not available & N/A \\
\hline
DC-AI-C3 & Text-to-Text Translation & 9.38\% & 6 \\
\hline
DC-AI-C4 & Image-to-Text & 23.53\% & 5 \\
\hline
DC-AI-C5 & Image-to-Image & Not available & N/A \\
\hline
DC-AI-C6 & Speech Recognition & 12.08\% & 4 \\
\hline
DC-AI-C7 & Face Embedding & 5.73\% & 8 \\
\hline
DC-AI-C8 & 3D Face Recognition & 38.46\% & 4 \\
\hline
DC-AI-C9 & Object Detection & 0 & 10 \\
\hline
DC-AI-C10 & Recommendation & 9.95\% & 5 \\
\hline
DC-AI-C11 & Video Prediction & 11.83\% & 4 \\
\hline
DC-AI-C12 & Image Compression & 22.49\% & 4 \\
\hline
DC-AI-C13 & 3D Object Reconstruction & 16.07\% & 4 \\
\hline
DC-AI-C14 & Text Summarization & 24.72\% & 5 \\
\hline
DC-AI-C15 & Spatial Transformer & 7.29\% & 4 \\
\hline
DC-AI-C16 & Learning to Rank & 1.90\% & 4 \\
\hline
DC-AI-C17 & Neural Architecture Search & 6.15\% & 6 \\
\hline
\end{tabular}
\end{table}

Without the equivalent benchmarking rules being explicitly stated, ImageNet/ResNet-50 training is not qualified for ranking the performance and energy efficiency of HPC AI systems. 

\section{Benchmarking Methodology}\label{sec:methodology}

This section presents our methodology to  achieve the goal of  
being equivalent, relevant, representative, affordable, and repeatable. 

\subsection{Equivalence}\label{subsec:equivalence}

To perform fair benchmarking across different systems or the same system with different scale,  we propose two approaches to assure the equivalence.  

First, as shown in Fig.~\ref{fig:nine-layer}, we abstract the system under test into nine independent layers, and put each layer under test while keeping the other layers intact unless  otherwise stated. 

Layer 1 is the hardware, including CPUs and networks. Layers 2, and 3 are the related system software, including the operating system (Layer 2), and  the communication libraries (Layer 3). Layer 4 is the AI accelerators, i.e., GPU,  and libraries, i.e., CUDA and cuDNN.  Layer 5 is the AI framework, such as TensorFlow~\cite{abadi2016tensorflow} and PyTorch~\cite{paszke2019pytorch}. Layer 6 refers to programming model, including parallel mode (data parallelism or model parallelism), and synchronous 
or asynchronous training. Layer 7 refers to the workloads used in HPC AI500 V2.0 benchmark. Layer 8 refers to hyper-parameters policies or settings. Layer 9 refers to problem domain, including datasets, target quality, and epochs.

Second, for the sake of simpleness, we propose three high levels of benchmarking and put several related layers together under test. 

(1) The hardware level. This high level is for benchmarking HPC AI hardware systems and their related system software (Layers 1, 2, 3, 4). In this context, the other layers should be kept intact unless otherwise stated in the benchmarking rules. The benchmark users should compile the source code of the benchmark implementation, provided by the benchmark committee, on their hardware directly with allowed changes. Luo et al.~\cite{luo2020comparison} show that the same model on different frameworks has different accuracy. So in addition to the same data set, and AI model, we mandate that the benchmark implementations also use the same AI framework. The benchmark users can  change hardware, OS, compiler settings, communication libraries. For the other layers, the benchmark users can only change parallel modes in Layer 6 or tune learning rate policies and batchsize settings in Layer 8.
It is the benchmark committee' duty to assure the equivalence of Layers 6, 7, 8, 9 across different benchmark implementations upon the users' requests. 


(2) The system level. Because of the portability cost, some benchmark users may opt for one specific AI framework without the support of the other, so specifying a fixed framework has a limited purpose. So in  the system level, we put the hardware system in addition to the AI framework  under the test (Layers 1, 2, 3, 4, and 5), which we call the system level. We mandate that the benchmark implementations use the same data set, and AI model.
In addition to the changes allowed in  the hardware level, the users are allowed to re-implement the algorithms on different or even customized AI framework (Layer 5).  The other layers should be kept intact unless otherwise stated in the benchmarking rules. 


The benchmark committee or  an independent group need double-check the  equivalence of Layers 6, 7, 8, 9 between the two benchmark implementations.

(3) The free level. In this high level, the specification of an AI task is stated in a paper-and-pencil manner separating from  its specific implementation. That is to say, the same data set, target quality, and training epochs  are defined in Layer 9 while the other layers are open for optimizations. 
The emphasis is advancing the state-of-the-art of software and hardware co-design, so the benchmark users can change any layers from Layer 1 to Layer 8 while keeping  Layer 9 intact. Meanwhile, the benchmark users are encouraged to disclose the details.


\subsection{Representative vs. Affordable}

We investigate and compare the state-of-the-art and state-of-the-practice of AI benchmark suites, including MLPerf~\cite{mattson2019mlperf}, AIBench~\cite{tang2020aibench}, Deep500~\cite{bhimji2018deep}, HPC AI500 V1.0~\cite{jiang2018hpc}. We present the detailed analytics in Section~\ref{related_work}. Fortunately, we found the methodology of AIBench and its subset combines the merits of two methodologies discussed in Section~\ref{challenge}. 

On one hand,  AIBench~\cite{tang2020aibench} is by far the most representative and comprehensive AI benchmark suite. 
It contains seventeen representative AI tasks.  These workloads are diverse in terms of model complexity, computational cost, and convergent rate, computation and memory access patterns, hotspot functions, and other micro-architecture characteristics. 

\begin{figure}[ht]
  \centering
  \includegraphics[width=\linewidth]{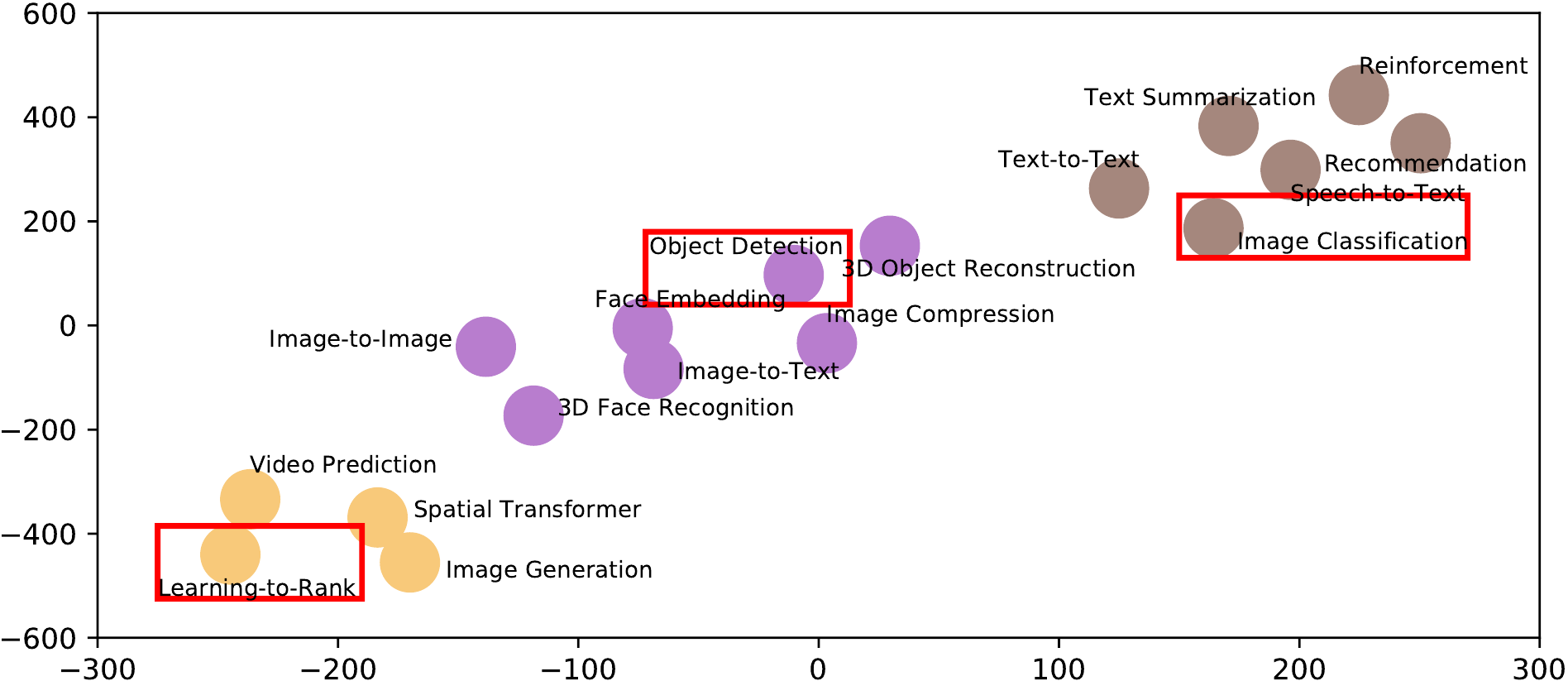}
  \caption{The three subset of AIBench with respect to the full benchmarks~\cite{tang2020aibench}. The clustering is based on the patterns of computation and memory access of seventeen AIBench component benchmarks, which described by five metrics listed in Table~\ref{table:clustering-metrics}. For visualization, five dimensional data are downscaled into two-dimension ones by the t-SNE clustering approach~\cite{tsne}.}
  \label{fig:aibench-clustering}
\end{figure}

On the other hand, for affordability, AIBench carefully selected a minimum subset from the seventeen AI tasks from perspectives of model complexity, computational cost, convergent rate, run-to-run variation, and having Widely accepted evaluation metrics or not. As shown in Fig.~\ref{fig:aibench-clustering}, the AIBench subset includes three AI tasks--Image Classification, Object Detection, and Learning to Rank. 

\begin{table}[ht]
\scriptsize
\centering
\caption{The metrics used by the t-SNE clustering approach~\cite{tang2020aibench}.}
\begin{tabular}[ht]{>{\centering}m{0.15\textwidth}|>{\centering\arraybackslash}m{0.7\textwidth}}
\toprule
\textbf{Metrics} & \textbf{Meaning} \\
\midrule
\textbf{achieved\_occupancy}& The ratio of the average active warps per active cycle to the maximum number of warps provided by a multiprocessor  \\
\midrule
\textbf{ipc efficiency} & The ratio of the executed instructions per cycle to the theoretical number.  \\
\midrule
\textbf{gld\_efficiency} & The ratio of the requested global memory load throughput to the required global memory load throughput  \\
\midrule
\textbf{gst\_efficiency} & The ratio of the requested global memory store throughput to the required global memory store throughput  \\
\midrule
\textbf{dram\_utilization} & The utilization level of the device memory relative to the peak utilization  \\

\bottomrule
\end{tabular}
\label{table:clustering-metrics}
\end{table}%

Tang et al.~\cite{tang2020aibench} systematically quantify the run-to-run variation of seventeen AI tasks of AIBench in terms of the standard deviation to the mean of the training epochs to achieve a convergent quality. The variation of image classification, object detection, and learning ranking is 1.12\%, 0\%, and 1.90\%, respectively, and they  are the  most  repeatable benchmarks, which is the other reason for including them into the subset.

So we choose the AIBench subset as the HPC AI500 V2.0 candidate benchmarks for implementing scalable HPC AI benchmark tools.

\subsection{Repeatability and Replicability}

In line with the experimental sciences discussed in ~\cite{Repli_reproducibility}, we propose the benchmarking procedures for assuring repeatability and replicability~\cite{Repro_vs_replica}. We adopt the definition similar to that of  the Association for Computing Machinery~\cite{ACM_defintion}. Different from reproducibility, which requires changes, repeatability and replicability avoid changes~\cite{Repli_reproducibility}.

Repeatability (same team): The benchmarking is performed on the same HPC AI system, using the same benchmark implementation under the same configurations, following the same benchmarking procedures,  on multiple trials~\cite{Repli_reproducibility}. 

The team should submit the raw data of all trials, including the average numbers in addition to its variations. 
The variation is measured in terms of the ratio of the standard deviation to the mean of the numbers of all trials.


To mitigate the influence of stochastic of the AI algorithm, each benchmark should mandate the least valid runs of benchmarking. The number of all trials should be more than the least valid runs of benchmarking. 

Replicability (Different team)~\cite{ACM_defintion}: The replicability refers to that the other team verifies the benchmarking results on the same HPC AI system, using the same benchmark implementation under the same configurations, following the same benchmarking procedures,  on multiple trials. 

For replicability, 
The benchmark committee or  an independent group need verify the numbers on the same system, and report the raw data of all trials, including the average numbers in addition to its variation.



\section{Benchmark Design and Implementation}\label{sec:desgin-implementation}

In this section, we firstly illustrate how to choose the workloads according to our benchmarking methodology (Section~\ref{sec:methodology}). Then we present the datasets, AI models, and reference implementations of HPC AI500. Finally, we introduce the metrics.

\subsection{How to choose the workloads?}

With respect to other AI benchmarks, there are two unique differences of HPC AI benchmarking. First, the challenges of HPC AI benchmarking inherit from the complexity of benchmarking scalable hardware and software systems at scale, i.e., tens of thousands of nodes, significantly different from that of IoT~\cite{luo2020comparison} or  datacenter~\cite{gao2020aibench}. On this point, we need consider the cost of benchmarking at scale. Second, HPC AI domains cover both commercial and high performance scientific computing. Currently, business applications are pervasive. Because of the difficulty of recruiting qualified scientists to label scientific data, 
the applications in scientific computing lag behind but are promising. In general, the scientific data are often more complex than that of the MINST or ImageNet data: the shape of scientific data can be 2D images or higher-dimension structures with  hundreds of channels, while the popular commercial image data like ImageNet often consist of only RGB~\cite{jiang2018hpc}. So we should include the scientific data in the HPC AI benchmarks. 




According to our benchmarking methodology discussed in Section~\ref{sec:methodology}, we choose the AIBench subset as the HPC AI500 candidate benchmarks for implementing scalable HPC AI benchmark tools.
 
\subsubsection{Representative and Affordable}
As the broad HPC AI applications cover both scientific~\cite{kurth2018exascale,kurth2017deep,mathuriya2018cosmoflow,racah2017extremeweather,liu2016application} and commercial field~\cite{tanakaimagenet,jia2018highly,ying2018image,akiba2017extremely}, we choose the most representative workloads and data sets from these two fields. 

\textbf{EWA} is one of the pioneering work that  uses deep learning algorithm to replace the rules predefined by human expert and achieve excellent results~\cite{liu2016application}. Most important of  all, the goal of EWA is to identify various extreme weather patterns (e.g. tropical depression), which is essentially ~\emph{object detection}--one of the three benchmarks of the AIBench subset. In 2018, a deep learning based  EWA implementation~\cite{kurth2018exascale} won the Gordon Bell Prize, which is the first AI application to win this award. 

\textbf{Image Classification} is widely used in many applications of \emph{commercial fields}, which is a fundamental task in AI research. With the developing of large-scale deep earning, Image Classification has become a well-known showcase optimizing HPC AI systems~\cite{tanakaimagenet,jia2018highly,ying2018image,akiba2017extremely}, as summarized in Table~\ref{table:imagenet-and-hpcai}. Image Classification is also one of the three benchmarks of the AIBench Subset. 

We exclude Learn to Ranking because it has the lowest computation complexity in terms of FLOPS, which is only 0.08 MFLOPs in terms of a single forward computation. According to~\cite{tang2020aibench}, Image Classification and Object Detection is more complex than that by one or two orders of magnitude, respectively. 

\subsubsection{Repeatability}

As the stochastic nature of AI, we need to ensure the repeatability by choosing relatively stable workloads in various AI tasks. According to the randomness analytics of AIBench~\cite{tang2020aibench}, the two most repeatable  AI benchmarks are Object Detection and Image Classification, whose variation is 0\% and 1.12\%, respectively. So they satisfies the property of a good benchmark--being repeatable.  

\subsubsection{Scaling Characteristics}
For comprehensive evaluation, the workloads we choose have distinct characteristics in terms of scaling characteristics. We use scaling ratio to depict the difficulty when scaling a workload from a single node to multiple nodes. As shown in Table~\ref{table:hpcaibench_scalingratio}, the scaling ratio of EWA and Image Classification is 16.85 and 117.76, respectively, reflecting very different scaling characteristics.  

\subsubsection{Different Levels of Stringent Quality Requirements}

When ranking HPC system, we consider not only its performance, but also the achieved quality. Different AI tasks have different levels of stringent quality requirement. 
Our benchmark decision also consider this factor. In our two benchmarks, EWA has much more stringent quality requirement than that of Image Classification.  

\subsection{Data sets and AI models}

\subsubsection{EWA}

\textbf{Dataset.} The EWA dataset~\cite{racah2017extremeweather} is made up of 26-year climate data. The data of every year is available as one HDF5 file. Each HDF5 file contains two data sets: images and boxes. The images data set has 1460 example  images (4 per day, 365 days per year) with 16 channels. Each channel is 768 * 1152 corresponding to one measurement per 25 square km on earth. The box dataset records the coordinates of the four kinds of extreme weather events in the corresponding images: tropical depression, tropical cyclone, extratropical cyclone and the atmospheric river.

\textbf{Model.} Faster-RCNN targets real-time Object Detection~\cite{ren2015faster}. As one of the latest models of an RCNN family~\cite{girshick2015fast,girshick2014rich}, it deprecates the selective search that has been used in the previous RCNN version. Instead, Faster-RCNN proposes a dedicated convolutional neural network, named region proposal network (RPN), to achieve nearly cost-free region proposals. With such design, Object Detection is much faster. As a result, Faster-RCNN wins the 1st-place entries in ILSVRC'15 (ImageNet Large Scale Visual Recognition Competition). 

\textbf{Quality}
The target quality is $MAP@[IoU=0.5]=0.35$, which is our best training result. MAP means the average precision, which is a dedicated metric for object detection. The IoU means the intersection over union, used to measure how much our predicted boundary overlaps with the ground truth. 

\subsubsection{Image Classification}

\textbf{Dataset.} ImageNet~\cite{deng2009imagenet} is large visual database designed for use in visual object recognition research. More than 14 million images have been hand-annotated according to the WordNet hierarchy. Both the original images and bounding boxes are  provided. The data size is more than 100 GB.

\textbf{Model.} ResNet is a milestone in Image Classification~\cite{he2016deep}, marking the ability of AI to identify images beyond humans in a particular domain. The spirit of ResNet is its success in reducing the negative impact of the degradation problem.  The degradation problem means in the very deep neural network, the gradient will gradually disappear in the process of back-propagation, leading to poor performance. Therefore, with ResNet, it is possible to build a deeper convolution neural network and archive the higher accuracy. Researchers successfully build a ResNet with 152 layers. This ultra-deep model won all the awards in ILSVRC'15. 

\textbf{Quality}
The target quality is $Top1$ $Accuracy = 0.763$, which is the highest accuracy by far in training ImageNet/ResNet50 at scale~\cite{ying2018image}. The Top-1 accuracy refers to that only the output with the highest probability is the correct answer.
\begin{table}[ht]
\scriptsize
\centering
\caption{The Summary of Image Data Sets of HPC AI500 V2.0 Benchmarks}
\begin{tabular}[ht]{>{\centering}p{0.25\textwidth}|>{\centering}p{0.2\textwidth}>{\centering}p{0.2\textwidth}>{\centering\arraybackslash}p{0.2\textwidth}}
\toprule
\textbf{Dataset}& \textbf{Channels} & \textbf{Resolution} & \textbf{Size}\\
\midrule
The extreme weather dataset~\cite{racah2017extremeweather}& 16 & 768*1052  & 558 GB   \\ \midrule
ImageNet dataset~\cite{deng2009imagenet} & 3 & 256*256  & 137 GB  \\ 

\bottomrule
\end{tabular}
\label{table:hpcaibench_dataset}
\end{table}%

\begin{table}[!ht]
\scriptsize
\centering
\caption{The scaling ratio of HPC AI500 v2.0 workloads}
\begin{tabular}[ht]{>{\centering}p{0.25\textwidth}|>{\centering}p{0.2\textwidth}>{\centering}p{0.2\textwidth}>{\centering\arraybackslash}p{0.2\textwidth}}
\toprule
\textbf{Workloads}& \textbf{Comm (Parameters/Step)} & \textbf{Comp (GFLOPs/Step)} & \textbf{Comp/Comm} \textbf{(GFLOPs/Parameters)}\\
\midrule
EWA & 41 million & 691  & 16.85   \\ \midrule
Image Classification & 25 million & 2944  & 117.76  \\ 

\bottomrule
\end{tabular}
\label{table:hpcaibench_scalingratio}
\end{table}%

\subsection{Reference Implementation}
The reference implementation of HPC AI500 V2.0 benchmark is summarized as shown in Table~\ref{table:benchmark_suite}. At present, we provide the implementations using TensorFlow~\cite{abadi2016tensorflow}, which is a popular deep learning framework in the HPC community~\cite{survey}. For communication, we adopt Horovod~\cite{sergeev2018horovod}
instead of the default GRPC protocol in TensorFlow, which is not extendable for large-scale cluster~\cite{mathuriya2017scaling} due to the limitation of the master-slave architecture and socket-based communication. Horovod is a library originally designed for scalable distributed deep learning using TensorFlow. It implements \textit{all\_reduce} operations using ring-based algorithms~\cite{ring-based} and other high efficient communication algorithms that are widely used in the traditional HPC community. 

\subsection{Metrics and Scoring Rules}\label{sec:metrics}

\subsubsection{Valid FLOPS}\label{sec:vflops}
We propose two metrics, called Valid FLOPS (in short VFLOPS) and Valid FLOPS per watt (in short VFLOPS per watt), to quantify the valid performance and energy efficiency that consider both the system throughput and model quality. The goal of these two metrics is to impose an penalty on failing to achieve a target quality.
VFLOPS and VFLOPS per watt is calculated according to the formulas as follows.

\begin{equation}
    VFLOPS=FLOPS*penalty\_coefficient
\end{equation}

The penalty\_coefficient is used to penalize or award the FLOPS if the achieved quality is lower or greater than the target quality. Its definition is described as follows:
\begin{equation}
\begin{split}
    penalty\_coefficient=(achieved\_quality/target\_quality)^{n}
\end{split}
\end{equation}

Here, $achieved\_quality$ represents the actual model quality achieved in the evaluation. $target\_quality$ is the state-of-the-art model quality that has been predefined in our benchmarks~\ref{table:benchmark_suite}. The value of n is a positive integer, which is used to define the sensitivity to the model quality. The higher the number of n, the more loss of quality drop. 
As EWA has much more stringent quality requirement than that of Image Classification.  We set n as 10 for EWA and 5 for Image Classification by default. 

We propose VFLOPS per watt to evaluate energy efficiency.



\begin{table}[ht]
\scriptsize
\centering
\caption{HPC AI500 V2.0 benchmark suite.}
 \begin{threeparttable}
\begin{tabular}[ht]{>{\centering}m{0.11\textwidth}|>{\centering}m{0.09\textwidth}>{\centering}m{0.13\textwidth}>{\centering}m{0.13\textwidth}>{\centering}m{0.13\textwidth}>{\centering}m{0.10\textwidth}>{\centering}m{0.10\textwidth}>{\centering\arraybackslash}m{0.09\textwidth}}
\toprule
\textbf{Problem Domains} & \textbf{Models} & \textbf{Datasets} & \textbf{Target Quality} & \textbf{AI Frameworks} &\textbf{Comm Lib\tnote{1}} &\textbf{AI Acc Lib\tnote{2}} & \textbf{Epochs}  \\
\midrule
EWA & FasterRCNN~\cite{ren2015faster}  & EWA~\cite{racah2017extremeweather}  & mAP@[IoU=0.5]=0.35 & TensorFlow & Horovod & CUDA, cuDNN, NCCL &50\\ \midrule
Image Classification & ResNet50 v1.5~\cite{he2016deep}  & ImageNet~\cite{deng2009imagenet} & TOP 1 Accuracy=0.763 & TensorFlow & Horovod & CUDA, cuDNN, NCCL & 90 \\ 

\bottomrule
\end{tabular}
\begin{tablenotes}
        \footnotesize
        \item[1] Comm Lib refers to the communication libraries. 
        \item[2] AI acc lib refers to AI accelerators libraries. 

      \end{tablenotes}
\end{threeparttable}
\label{table:benchmark_suite}
\end{table}%

\section{Benchmarking Rules and Procedures} ~\label{sec:benchmarking-rules}

For the fairness and equivalence of benchmarking different HPC AI systems, a series of clear and unambiguous benchmarking rules are mandatory. 

Our fundamental benchmarking rule is that we put each independent layer (Shown in Fig.~\ref{fig:nine-layer}) under test while keeping the other layers intact.

Furthermore, for the hardware-level and system-level benchmarking presented in Section~\ref{sec:methodology}, we give a detailed description from perspectives of each layer. Finally, we introduce the benchmarking procedures.

\subsection{The Benchmarking Rules for the Hardware Level}


Based on our nine-layer model (Fig.~\ref{fig:nine-layer}), we specify the rules of each layer from top to bottom. 

\subsubsection{Problem Domain Layer}
\begin{itemize}
    \item The dataset and target quality must be in accordance with the specification of HPC AI500 V2.0 benchmark that we have discussed in Section~\ref{sec:desgin-implementation}. 
    \item The training epoch number should be the same like the reference implementation to guarantee the equivalent computational cost, namely 90 epochs for ImageNet and 50 epochs for EWA. Note that an epoch is an iteration over the entire data set, while a step refers to one update of the model parameters. The number of epochs is based on our experimental observation, and it should be updated in the future as well as the target qualities. 
\end{itemize}

\subsubsection{Hyper-parameters Setting Layer}
The rules of hyper-parameters setting layer include three parts, namely batchsize setting, learning rate policies, and other hyper-parameters settings. 

\paragraph{Batchsize Setting}
The batchsize of a training step is allowed to change, to fully utilize the computing capability of the system. 

\begin{figure}[h]
\begin{subfigure}{0.48\textwidth}
  \centering
  \includegraphics[width=.9\linewidth]{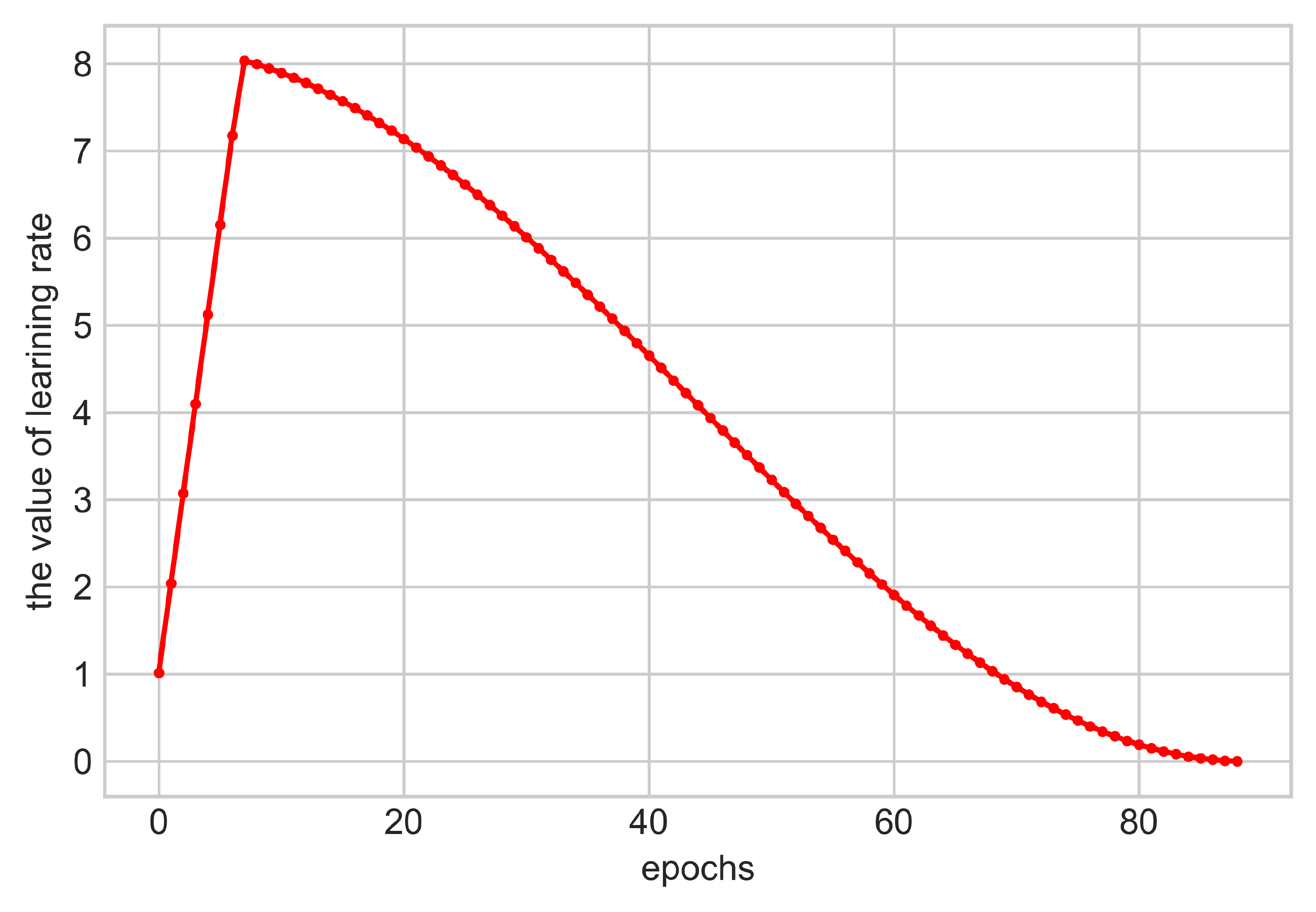}
  \caption{The learning rate curve.}
  \label{fig:learning-rate-curve}
\end{subfigure}
\begin{subfigure}{0.48\textwidth}
  \centering
  \includegraphics[width=.9\linewidth]{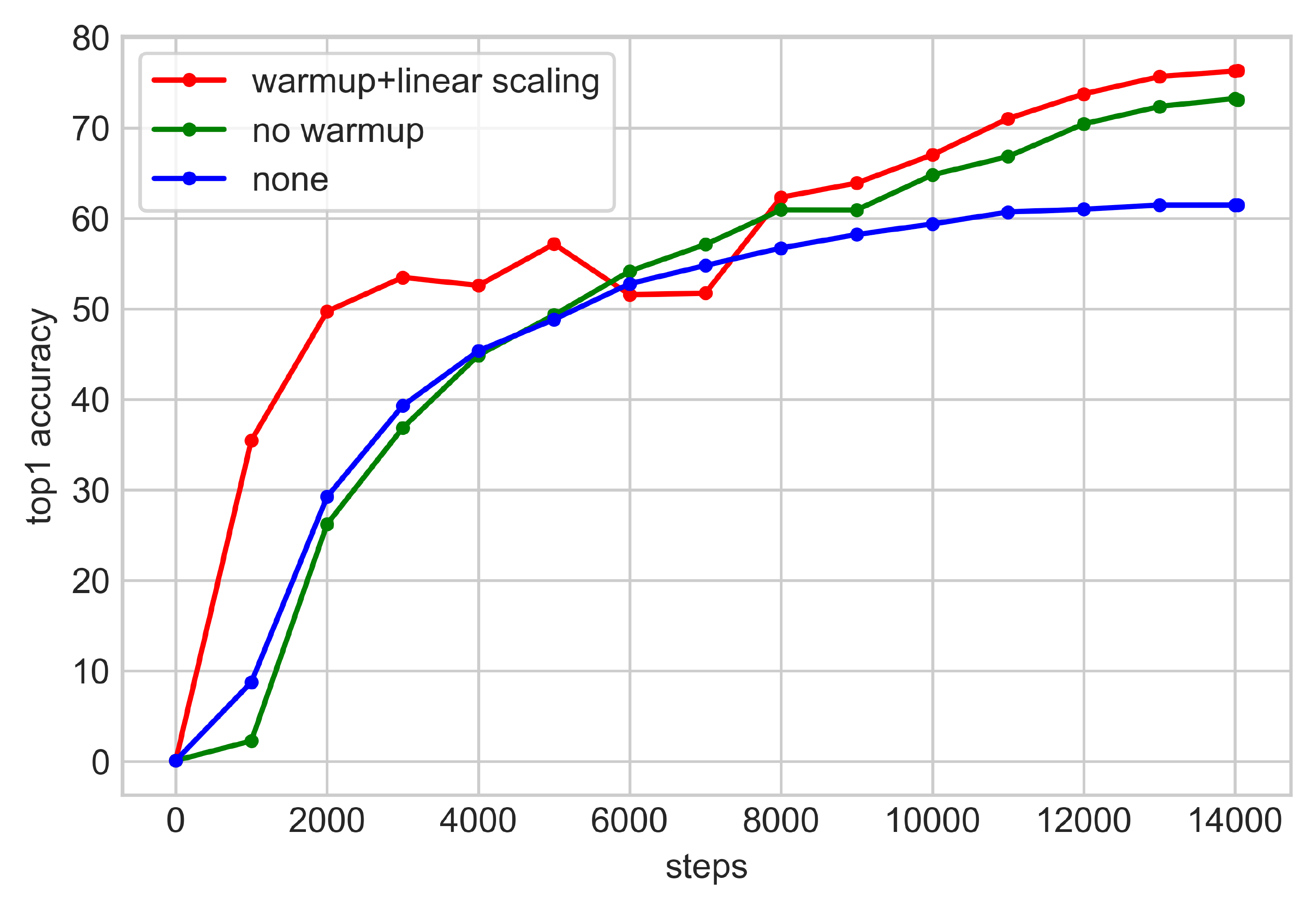}
  \caption{The effect on accuracy.}
  \label{fig:lr-and-warmup-effect}
\end{subfigure}

\caption{The learning rate curve and its effect on accuracy with linear scaling and warm-up rules. The benchmark is Image Classification and the system scale is 64 GPUs. The experiment configuration is consistent with that of Table~\ref{table:hwconfigeration}. The batchsize is 8192. }
\label{fig:warmup_effect}
\end{figure}

\paragraph{Learning Rate Policies}
Previous work shows the increase of batchsize leads to a fall of the model quality~\cite{smith2017don}. In this context, many learning rate policies are proposed~\cite{krizhevsky2014one, you2017imagenet, goyal2017accurate, you2019large}. With state-of-the-art learning rate policies, we can increase the  training batch size to fully utilize the hardware's resources while preserving the model quality at the same time. As each learning rate policy has its limitation in terms of the maximum supporting batchsize, our rule allows benchmark users to propose new learning rate policies to fully utilize the hardware's resources. Meanwhile, we provide a default learning rate policy.
\subparagraph{The default learning rate policy:}
The default learning rate policy of HPC AI500 is a linear scaling rule and a warm-up rule. The description is as follows:
\begin{itemize}
    \item A linear scaling rule: multiply the base learning rate $\eta$ by $k$ when the batch size is multiplied by $k$. The goal of the linear scaling rule is to make SGD updates similar in both distributed and single-worker training~\cite{goyal2017accurate}.
    \item A warm-up rule: gradually increase the learning rate from a small to a large number until it equals to $\eta\times k$. After warmup, the learning rate starts the original learning rate schedule (e.g. cosine decay). The Warm up rule is proposed since using linear scaling rule alone breaks down when the weight of the neural network is changing rapidly in the early training stage~\cite{goyal2017accurate}. 
  
Fig.~\ref{fig:learning-rate-curve} shows the learning rate changing curve after applying linear scaling and warm-up rules. We also perform a series of experiments to show the effect of this policy on model quality. As shown in Fig.~\ref{fig:lr-and-warmup-effect}, linear scaling and warm-up rules can improves the top1 accuracy from 61.48\% to 76.34\% in Image Classification when the batchsize is 8192. 

\end{itemize}

\subparagraph{Other learning rate policies:} Except for the linear scaling and warmup scheme, using state-of-the-art learning rate policies (e.g. LARS~\cite{you2017imagenet} and LAMB~\cite{you2019large}) are allowed. For new proposed ones, benchmarking users should open source their methods. 

\paragraph{Other Hyper-parameters Setting}
Except for batchsize and learning rate policy, other hyper-parameters such as weight decay, momentum must be as the same as the reference implementation.

\subsubsection{Workload Layer}
The AI algorithms in the workload must be the same as the reference implementation. 

\subsubsection{Programming Model Layer}
\begin{itemize}
    \item Data parallelism and model parallelism are both allowed as long as the mathematical equivalence is preserved. 
    \item Synchronous  Stochastic Gradient Descent (SGD) must be used in training, since asynchronous SGD may a) introduce the randomness, b) destroy the mathematical equivalence, c) decrease the accuracy.
\end{itemize}

\subsubsection{AI Framework Layer}
The AI framework must be the same as the reference implementation.

\subsubsection{Communication Libraries Layer}
In synchronous communication, the workers in the cluster 
must wait until all the workers have finished, to proceed to next iteration. We allow different communication policies in an  synchronous mode. 


\begin{table}[ht]
\scriptsize
\centering
\caption{Some common communication typologies of allreduce.}
\begin{tabular}[ht]{>{\centering}m{0.33\textwidth}|>{\centering\arraybackslash}m{0.48\textwidth}}
\toprule
\textbf{Topologies} & \textbf{Applications} \\
\midrule
Butterfly & OpenMPI~\cite{openmpi}\\
\midrule
Double binary tree & NCCL~\cite{NCCL} \\
\midrule
Ring & Baidu DeepSpeech~\cite{hannun2014deep}, Horovod~\cite{sergeev2018horovod} \\
\midrule
Hierarchical ring &	Horovod  \\

\bottomrule
\end{tabular}
\label{table:allreduce}
\end{table}%

\begin{itemize}
\item Based on AllReduce. Table~\ref{table:allreduce} shows the common topology used in AllReduce implementations. The benchmark users are allowed to utilize these existing ones or propose new typologies according to the configuration of the systems. For example, the researchers from Lawrence Berkeley National Laboratory archived Exascale FLOPS by customizing a communication topology of AllReduce on SUMMIT~\cite{kurth2018exascale}.

\item Based on MapReduce. The communication topology is determined by the implementation of MapReduce. The distributed training of Spark MLlib, SystemML, and REEF are all based on MapReduce. Users are allowed to implement customized MapReduce on their systems.

\item Based on parameter server. It is mandatory that only the synchronous mode is used for the parameter server, while it  also supports  asynchronous training.
\end{itemize}
\subsubsection{The AI Accelerators and Libraries Layer}
\begin{itemize}
\item Benchmark users can choose the AI accelerator library to achieve the best performance out of the system.
\item The single-precision floating point (FP32), half precision (FP16, BFLOAT16~\cite{googlebfloat}), and quantization (INT8, INT4) are allowed.
\end{itemize}

\subsubsection{OS Layer}
\begin{itemize}
\item Benchmark users can adjust OS configurations (such as CPU-Affinity setting) to achieve the best performance out of the system.
\item Benchmark users can choose `-O2' compiler optimization option when compiling the benchmarks and the run time environment software.
\end{itemize}

\subsubsection{Hardware Layer}

Benchmark users can adjust hardware configurations (such as hyper-threading setting, memory-prefetching setting) to achieve the best performance out of the system. 


\subsection{The Rules for The System Level}

As discussed in Section~\ref{subsec:equivalence}, in  the system level, we put the hardware system in addition to the framework under test. Therefore, in addition to the rules defined in the hardware level, benchmark users are allowed to reimplement the benchmark using a different or even customized AI framework at the AI framework layer. 

\subsection{The Benchmarking Procedures}
\subsubsection{Deployment}
Benchmark users need to download the source code of the benchmarks from the Benchcouncil Web site.
\subsubsection{Measurement}
\begin{itemize}
    \item Timing rules: timing starts when the workload reads the first batch training data and ends when the target epochs is reached. 
	\item Runs: according to the variation of EWA and Image Classification from Table~\ref{fig:AIBench-randomness}, the least number of runs is 5 and 10, respectively, to reduce run-to-run variation. 
	For reporting, we drops the runs with the highest and lowest variations, than calculate the arithmetic mean of the remaining results. 
	\item Benchmarking scores: 
	
	1) time-to-quality is the training time to its achieved quality;
	
	2) FLOPS refers to the single-precision floating point operations (or equivalent operations) per second.  The equivalent operations of the single-precision floating point operations include but not limited with FP16, BFLOAT16, INT8, and INT4; 
	
	3) VFLOPS and VFLOPS per Watt refers to the definitions in Section~\ref{sec:vflops}.
\end{itemize}

\subsubsection{Reporting Procedure}
The reporting results should include the following parts:
\begin{itemize}
	\item The description of system under test, including but not limited to: 
	
	1) detail descriptions of parameters of CPUs and AI accelerators in a single-node; 
	
	2) detail descriptions of parameters of intra-node connection in a single-node; 
	
	3) detail descriptions of parameters of OS in a single-node; 
	
	4) detail descriptions of parameters of run time environment software in a single-node; 
	
	5) detail descriptions of parameters of inter-node connection in the system; 
	
	6) detail descriptions of parameters of run time environment software in the system.
	
	\item Benchmark configurations, including but not limited to: 
	
	1) all hyper-parameter setting; 
	
	2) detail descriptions of communication.
	
	\item Benchmarking scores, including time-to-quality, FLOPS, FLOPS per Watt,  VFLOPS and VFLOPS per watt in all runs. These metrics should be submitted with the output log of the benchmark. 
	\item The source code, relevant document, and running script should be uploaded to Benchhub, which is the official code repository managed by BenchCouncil.
\end{itemize}
The BenchCouncil community is responsible for checking the replicability of the reported results and reviewing the code.

\subsection{Why Equivalent Benchmarking Rules Matter?} \label{sebsec:evaluation-equivalence}

A lot  of previous work~\cite{tanakaimagenet,jia2018highly,ying2018image,akiba2017extremely} focuses on accelerating Image Classification/ResNet-50 training. These efforts reduce the training time from hours to minutes. In this section, we take Image Classification as an example to explain why equivalent benchmarking rules matter for fair ranking HPC AI systems.

Batch normalization is a common effective method to improve the model generalization~\cite{ioffe2015batch}. The trainable parameters of batch normalization $\gamma$ and $\beta$ are used to restore the representation ability of the network. Jia et al. ~\cite{jia2018highly} propose eliminating  the weight decay on $\gamma$ and $\beta$ of batch normalization layer, which is a significant algorithm innovation in their work.  We re-implement this algorithm-level optimization in accordance with~\cite{jia2018highly}. Further, we use the VFLOPS as the metric to quantify the performance gap. 

The benchmarking results are  shown in Table~\ref{table:bn_no_wd}. The accuracy gain and corresponding VFLOPS ratio  are reported against the one without removing the weight decay. 
We find that as the system scale becomes larger, this optimization has a greater impact on the achieved quality. The accuracy gain is 0.45\% on the scales of 16 and 32 GPUs, and then jumps to 1.38\% on the scale of 64 GPUs, which is a notable improvement.  We calculate the VFLOPS ratio according to the formula discussed in Sec~\ref{sec:metrics} for each system scale. On the system scale of 64 GPUs, the VFLOPS ratio is high as 1.10, which is essentially the gain contributed solely  by the algorithm innovation. 

Consider the following case: we perform a comparison between two HPC AI systems using the same benchmark. One benchmark user leverages this  algorithm innovation, while the other does not. If we do not exclude this case in the benchmarking rules, the benchmarking results will be unfair. That is the reason why we mandate that the other hyper-parameter settings in Layer 8 must keep intact  as shown in Fig.~\ref{fig:nine-layer}.

Someone may question why we allow changing learning rate policies in Layer 8 in our rules as shown in Fig.~\ref{fig:nine-layer}. Just as discussed in Section~\ref{sec:benchmarking-rules}, this is because to fully utilize the hardware resources, the users have to change the learning rate policies.   

\begin{table}[ht]
\scriptsize
\centering
\caption{The impact of removing the weight decay on batch normalization (BN) layer with different system scales. The benchmark is Image Classification and the accuracy is measured by Top-1 accuracy.}
\begin{threeparttable}
\begin{tabular}[ht]{>{\centering}m{0.2\textwidth}|>{\centering}m{0.2\textwidth}|>{\centering}m{0.2\textwidth}|>{\centering\arraybackslash}m{0.2\textwidth}}
\toprule
\textbf{System Scale}&\textbf{Batchsize} & \textbf{Accuracy Gain}  & \textbf{VFLOPS Ratio\tnote{1}}\\
\midrule
16 GPUs & 2048 & +0.45\%  & 1.03 \\ \midrule
32 GPUs & 4096 & +0.45\%  & 1.03  \\ \midrule 
64 GPUs & 8192 & +1.38\%  & 1.10  \\ 
\bottomrule
\end{tabular}
\begin{tablenotes}
        \footnotesize
        \item[1] The VFLOPS ratio refers to the ratio of the VFLOPS after the optimization against the one without optimization.

\end{tablenotes}
\end{threeparttable}

\label{table:bn_no_wd}
\end{table}

\section{The HPC AI Roofline Performance Model}
Given a specific HPC AI system, the theoretical peak performance number can be calculated according to hardware configurations. However, the theoretical peak one is hard to achieve. Hence, we need a performance model to help achieve the  upper bound performance of an HPC AI system.

The previous Roofline model~\cite{williams2009roofline} is a upper bound performance model based on FLOPS and operation intensity (OI)--the total number of floating point instructions divided by the total byte number of memory accesses. With the aid of a Roofline model, we can decide a workload is memory-bound or compute-bound. Moreover, potential optimization strategies can be recommended according to the different ceilings of the Roofline model. 
To date, there is no such a performance model available for HPC-AI systems. 
In this section, we first analyze the distinctive characteristics of an HPC-AI system, and then propose an HPC-AI Roofline Model.



\subsection{The Architecture of  an HPC AI System} 
\begin{figure}[ht]
  \centering
  \includegraphics[width=\linewidth]{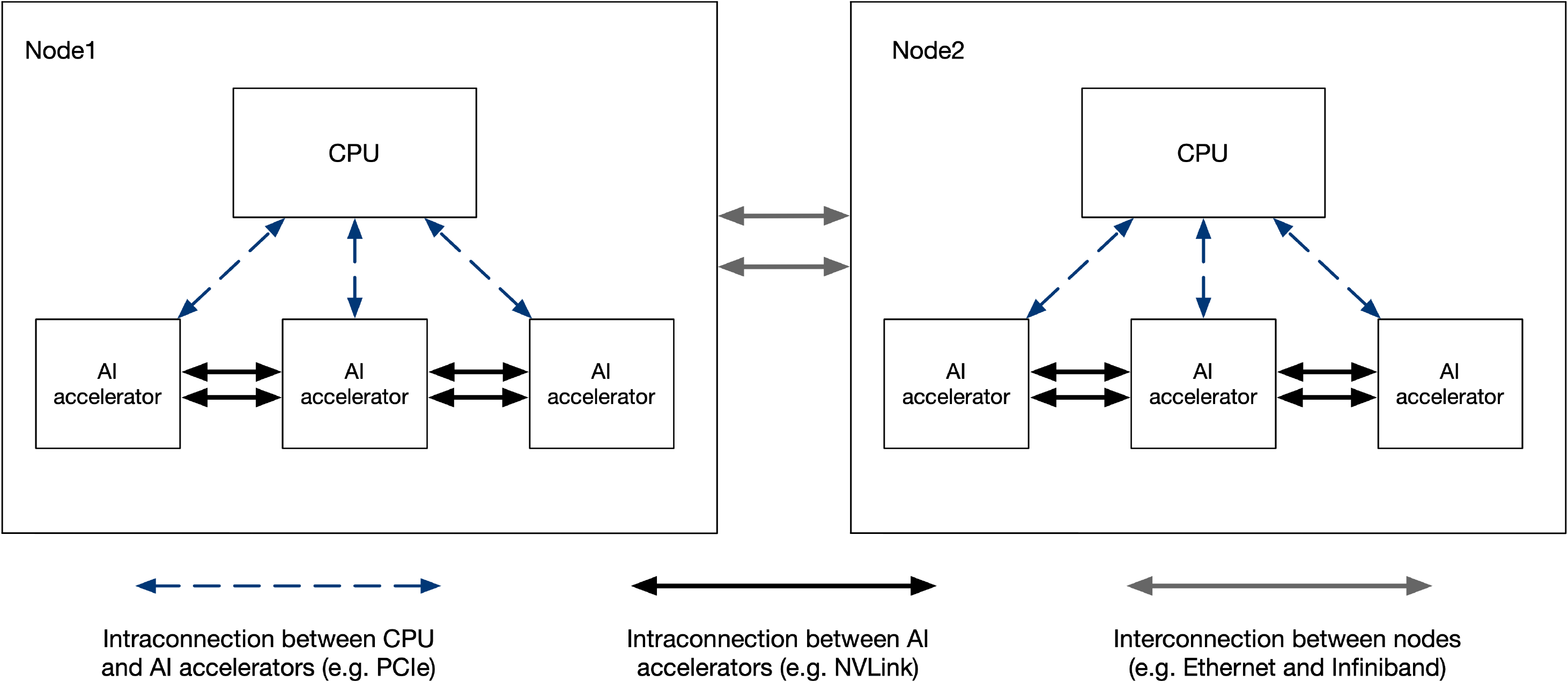}
  \caption{The Architecture of an HPC AI system.} 
  \label{fig:hpcai-systems-schematic-draw}
\end{figure}

An HPC AI system is a distributed system consisting of multiple nodes, each of which is heterogeneous and equipped with multiple CPUs and AI accelerators, as shown in Fig~\ref{fig:hpcai-systems-schematic-draw}.
The CPUs of each node are responsible for scheduling tasks and communicating with other nodes. The AI accelerators are responsible for AI calculations. Each AI accelerator
loads or stores data from its memory units through memory channels. And all AI accelerators of each node are connected with a specific high-speed network (e.g. NVLink for GPUs). 
The distributed nodes are interconnected by a general high-speed network (e.g. high speed Ethernet). 
Hence, the communications include both inter-node and intra-node ones. Our analytics in Section~\ref{subsec:scaling-expriments} reveals the communication efficiency is one of the dominant factors that impact its performance.


When proposing HPC-AI Roofline models, we consider the distinctive characteristics of HPC AI systems and the huge impact of communication efficiency on the performance of HPC AI systems. Significantly different from the original Roofline model~\cite{williams2009roofline}, which emphasizes the impact of  computation (FLOPS) and memory access (OI) on the overall performance, our HPC-AI Roofline model emphasizes the impact of communication and computation. 
We propose an innovative metric, named communication operation intensity (in short, COI), to replace OI. COI  is  defined as the total number of floating point instructions divided by the total byte number of communication.


Considering the different communication modes of inter-node communication (general high speed network) and intra-node communication (specific high speed network), our HPC-AI Roofline model is a combination of a single-node  model with a distributed  model. 

We use FLOPS as the metric to depict the upper bound performance. Unlike the original Roofline model~\cite{williams2009roofline} using the double-precision floating point operations per second, we use the single-precision floating point operations or equivalent operations, such as mixed-precision floating point operations per second. This is because double-precision floating point operations are rarely required for deep learning workloads,  while single-precision or mixed-precision floating point operations are prevalent.  

Intentionally, we do not choose VFLOP as the performance metric. This is because the purpose of the Roofline model is to decide the performance bound of the workload and guide its system-level and hardware-level optimizations. Instead, VFLOP is a composite metric reflecting both performance and accuracy to rank the HPC AI systems. 


\subsection{The Single-Node HPC-AI Roofline Model}\label{single_node_roofline}

The single-node HPC-AI model is formulated as follows.
\begin{equation}
\begin{split}
FLOPS_{Attained}=min(FLOPS_{Peak}, ComBand_{Peak}*COI)
\end{split}	
\end{equation}
$ComBand_{Peak}$ is the theoretical peak communication bandwidth of a single-node HPC AI system, which is the bandwidth of interconnections among AI accelerators. $FLOPS_{Peak}$ is the theoretical peak FLOPS of a single-node HPC AI system,  which is the aggregate theoretical peak FLOPS of all AI accelerators. The communication operation intensity--$COI$--is obtained by $COI=FLOPs/CT$ where $CT$ is short for the communication traffic--the total number of communication bytes among AI accelerators. To more accurately reflect the performance bottleneck of a given workload, different ceilings are added to help locate the bottlenecks and provide potential optimization 
recommendations. 

We use CONV (convolution) and GEMM (GEneral Matrix to Matrix Multiplication) to measure the upper bound performance of the system. On one hand, they are two most frequently-appearing kernel functions of the seventeen benchmarks of AIBench; On the other hand, their computing patterns, i.e., their multiplying and adding calculations can be fused, allow them to make more efficient use of accelerators. $FLOPS_{Attained}$ is the performance that a workload can attain, and the attained performance bound of a given workload under ceilings is formulated as follows.
\begin{equation}
\begin{split}
FLOPS_{Attained}=Min(FLOPS_{Ceiling}, ComBand_{Ceiling}*COI)
\end{split}	
\end{equation}

\subsection{The Distributed HPC-AI Roofline Model}

For the distributed model, we propose using COI (communication operation intensity) and FLOPS to depict the upper bound performance.
The model is formulated as follows.
\begin{equation}
\begin{split}
FLOPS_{Attained}=Min(FLOPS_{Peak}, ComBand_{Peak}*COI)
\end{split}	
\end{equation}
The $ComBand_{Peak}$ is the theoretical peak communication bandwidth of the distributed system, i.e., the theoretical bandwidth of the high speed Ethernet. $FLOPS_{Peak}$ is the theoretical peak FLOPS of the distributed system,  which is the aggregate theoretical FLOPS of all AI accelerators in the distributed system. The communication operation intensity--$COI$ is obtained by $COI=FLOPs/CT$, where the communication traffic--$CT$ is the total byte number of communications among all AI accelerators in the distributed system. To more accurately reflect the performance bottleneck of a given workload, we  add several ceilings,  and the attained performance bound of a given workload  is formulated as follows.
\begin{equation}
\begin{split}
FLOPS_{Attained}=Min(FLOPS_{Ceiling}, ComBand_{Ceiling}*COI)
\end{split}	
\end{equation}

\begin{figure*}[ht]
\begin{subfigure}{0.48\textwidth}
  \centering
  \includegraphics[width=.9\linewidth]{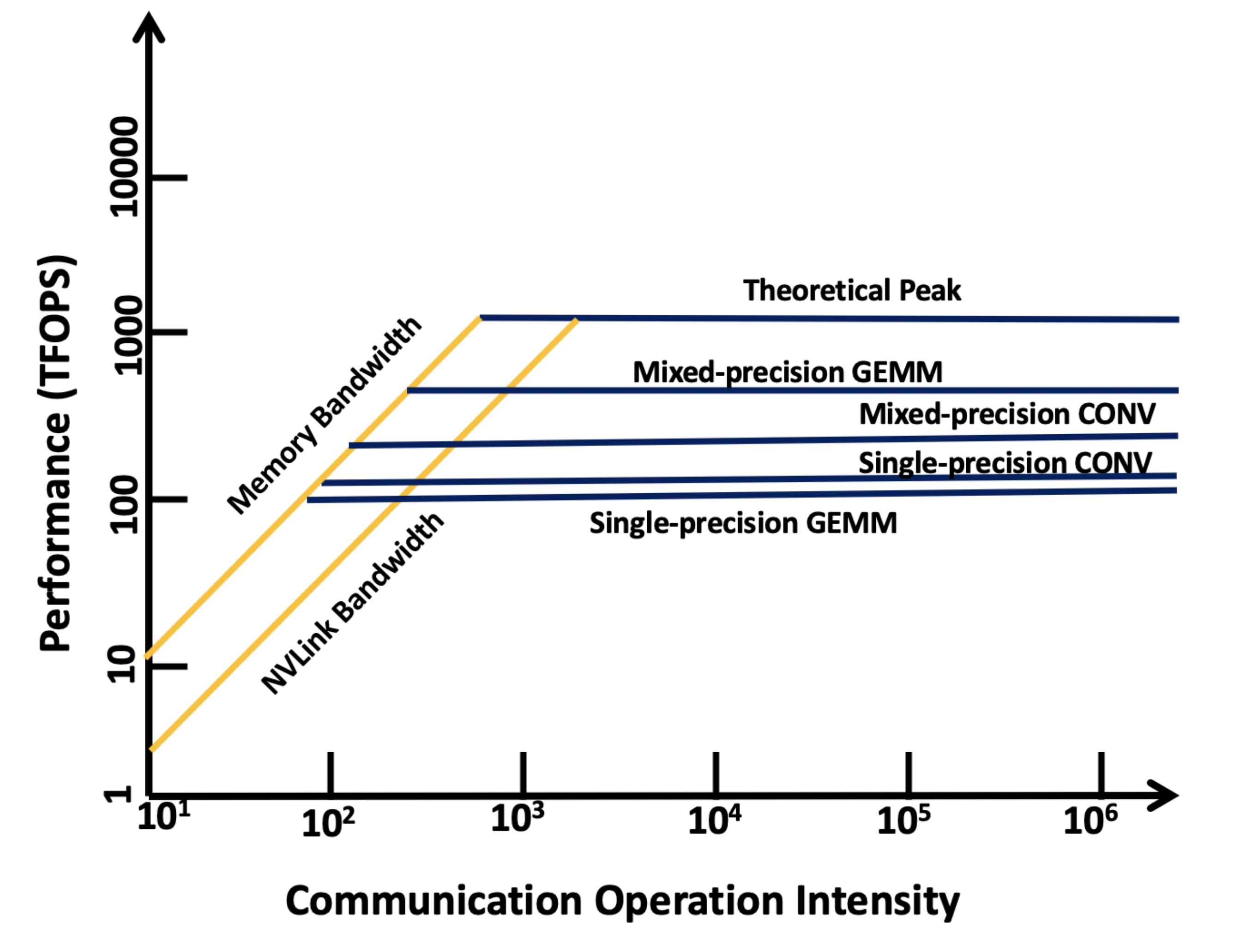}
   \caption{The Single-Node version. }
  \label{fig:SingleNodeRoof}
\end{subfigure}
\begin{subfigure}{0.48\textwidth}
  \centering
  \includegraphics[width=.9\linewidth, height=5.2cm]{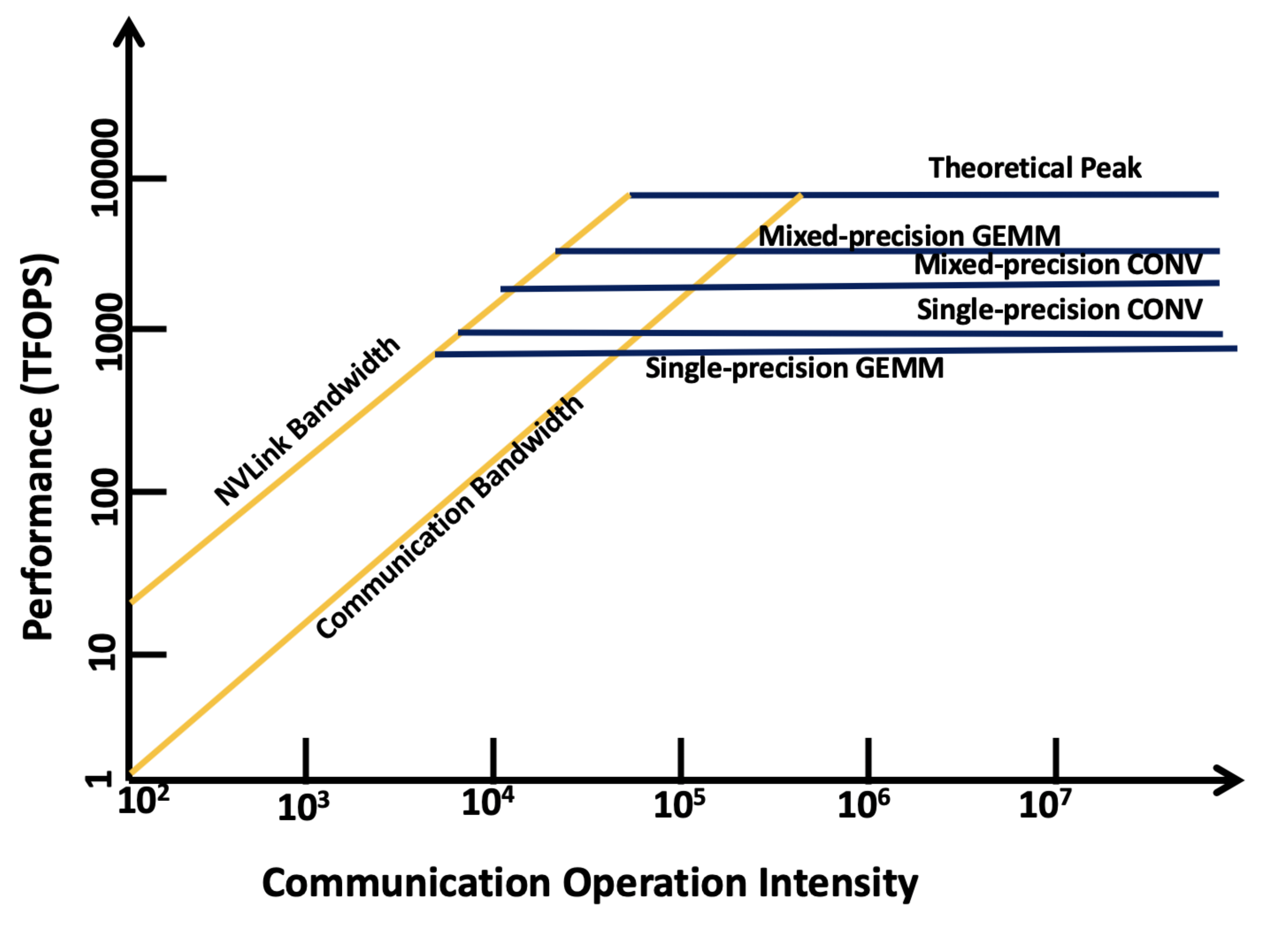}
  \caption{The Distributed version.}
  \label{fig:DisNodeRoof}
\end{subfigure}
\caption{The HPC-AI Roofline Model.}
\end{figure*}

\subsection{A Case Study of the HPC AI Roofline Models}

We perform a case study of our HPC AI Roofline models on an experimental system. The system consists of eight nodes, each of which is equipped with one Intel(R) Xeon(R) Platinum 8268 CPU and eight NVIDIA Tesla V100 GPUs. Each GPU in the same node has 32 GB HBM memory, connected by NVIDIA NVLink–a high-speed GPU interconnection that has theoretical peak 300GB/s bi-directional bandwidth. The nodes are connected with an Ethernet networking with a bandwidth of 10 Gb/s. Each node has 1.5 TB of system memory and 8 TB of NVMe SSD disk.

\subsubsection{A Case Study of the Single-Node HPC-AI Roofline Model}

As shown in Fig.~\ref{fig:SingleNodeRoof}, the y-axis is the performance in terms of floating-point operations per second, while the x-axis is the communication operation intensity--the floating-point operations divided by the total byte number of communication. In Fig.~\ref{fig:SingleNodeRoof}, the peak computation rate forms the `flat' part, while the communication bandwidth turns into the `slanted' part. So, if the communication operation intensity is lower, the workload is communication-bound, under the slanted part of the roofline. With the sufficient communication operation intensity, the workload is compute-bound.

We add four computation ceilings: mixed-precision GEMM (the performance of the mixed-precision floating point implementation of GEMM), single-precision GEMM, mixed-precision CONV, and single-precision CONV.  Single-precision setting is commonly-used in the AI domain, while mixed-precision is one of the optimization features on some advanced AI accelerators. 

The best-case performance of eight GPUs is that the communication and computation totally overlap, and the memory bandwidth becomes the bottlenecks.
We add one communication ceiling -- memory bandwidth.  In Fig.~\ref{fig:SingleNodeRoof}, the theoretical peak number of mixed-precision FLOPS, the mixed-precision GEMM ceiling, the mixed-precision CONV ceiling, the single-precision GEMM ceiling, the single-precision CONV ceiling is 1040  TFLOPS,  636 TFLOPS, 176 TFLOPS,  115 TFLOPS, 112 TFLOPS, respectively. Note that the gap between the theoretical perk number with the actual one is because that the performance of CONV and GEMM is affected by the dimension and sparsity of input data, NCHW format and output channels. Additionally, the convolution kernel also impacts the performance of CONV greatly. 
The different input  size of CONV and GEMM leads to different performance numbers. The NVLink ceiling is the theoretical peak bandwidth of the communications among GPUs--300 GB/S, and the memory bandwidth ceiling is the theoretical peak bandwidth of the memory--1134 GB/S.

\subsubsection{A Case Study of the Distributed HPC-AI Roofline Model}
Our system consists of eight nodes. All the GPUs in the same node are connected by NVIDIA NVLink, and the nodes are connected with an Ethernet networking. In Fig.~\ref{fig:DisNodeRoof}, the peak computation rate forms the `flat' part, while the communication bandwidth (Ethernet networking bandwidth) turns into the `slanted' part. The theoretical Peak FLOPS of the system is 8320 TFLOPS, and the communication ceiling is 1.2 GB/S.  

We add four computation ceilings: mixed-precision GEMM, single-precision GEMM, mixed-precision CONV, and single-precision CONV. Their numbers are 5091, 920, 2376, and 976 TFLOPS, respectively.  

The best-case performance of the HPC-AI system is that the communications are within the nodes. So we add one communication ceilings--NVLink bandwidth. The NVLink bandwidth ceiling is 300 GB/S.


\begin{table}
\caption{Hardware configuration details.}\label{table:hwconfigeration}
\renewcommand\arraystretch{1.1}
\center
\scriptsize
\begin{tabular}{>{\centering}m{0.18\textwidth}|>{\centering}m{0.15\textwidth}?>{\centering}m{0.18\textwidth}|>{\centering\arraybackslash}m{0.2\textwidth}}

\hline \rowcolor{mygray} \multicolumn{2}{c}{System Configurations}&\multicolumn{2}{c}{Single-Node Configurations}\\
\hline Num of Nodes & 8 & CPU Type & Intel(R) Xeon(R) Platinum 8268 CPU \\
\hline GPUs per Node & 8 & Memory & 1.5TB, DDR4 \\
\hline Total num of GPUs & 64 & Disk & 8TB, NVxMe SSD \\
\hline Peak Theoretical performance (FP32) & 960 TFLOPS & GPU Type & Nvidia Tesla V100 \\
\hline Peak Theoretical performance (Mixed) & 7680 TFLOPS & GPU Memory & 32GB, HBM \\
\hline Interconnection & Ethernet, 10Gb/s & Intraconnection & NVLink \\
\hline

\end{tabular}
\end{table}
\section{Evaluation}
In this section, we introduce the experimental configurations in Section~\ref{subsec:ex-config}, present how to measure FLOPs in Section~\ref{subsec:calculate-flops}.
Then, we perform an in-depth performance analysis of a single node in Section~\ref{subsec:single-node} and multiple nodes in Section~\ref{subsec:scaling-expriments}, respectively. Finally, we demonstrate how to use our roofline model to guide the optimizations of the HPC AI systems in Section~\ref{subsec:roofline-model}.

\subsection{Experimental Configurations}\label{subsec:ex-config}

Our experiments are conducted on an HPC AI system, consisting of eight nodes, each of which is equipped with one Intel(R) Xeon(R) Platinum 8268 CPU and eight NVIDIA Tesla V100 GPUs. Each GPU in the same node has 32GB HBM memory, connected by NVIDIA NVLink--a high-speed GPU interconnection whose theoretical peak bi-directional bandwidth is 300GB/s. The nodes are connected with an  Ethernet networking with a bandwidth of 10 Gb/s. Each node has 1.5 TB  system memory and 8 TB NVMe SSD disk. 

The details of the architecture of  each NVIDIA Tesla V100 GPU--NVIDIA Volta architecture are as follows.
The NVIDIA Volta architecture is equipped with 640 Tensor Cores to accelerate GEMM and convolution operations.
Each Tensor Core performs 64 floating-point fused-multiply-add (FMA) operations per clock, delivering up to 125 TFLOPs of theoretical peak performance. When performing mixed precision training with a Tensor Core, we uses FP16 for calculation and FP32 for accumulation~\cite{micikevicius2017mixed}. 

We use TensorFlow v1.14, compiled with CUDA v10.1 and cuDnn v7.6.2 backend. We use Horovod v0.16.4 for synchronous distributed training, compiled with OpenMPI v3.1.4 and NCCL v2.4.8. 
NCCL is short for the NVIDIA Collective Communications Library, which is  a closed-source library of multi-GPU collective communication primitives that are topology-aware.

\subsection{Performance Measurement} \label{subsec:calculate-flops}



The source-code level measurement of FLOPs is difficult for a complex AI model implemented with a complex AI framework. The mainstream frameworks like TensorFlow and PyTorch adopt computational graphs and map them to specific computing engines, e.g., GPU and cuDNN. This process invokes numerous kernels, and each of which contributes to a portion of FLOPs. Hence, we need to figure out the implementation of each invoked kernel to obtain the FLOPs of an entire AI model. Unfortunately, the source code is not publicly available as the NVIDIA libraries, like CUDA and cuDnn are not open source.


We use NVProf~\cite{nvprof}--a performance analysis tool for NVIDIA GPUs--to measure the FLOPs in our experiments. NVProf can be used to collect the profiling data from hardware performance counters. But it  has a huge overhead, slowing down the the execution time more than hundreds of times. Thus, profiling the whole training session of a deep learning model is prohibitively costly.
The previous work~\cite{gao2018bigdatabench,zhu2018tbd} has found that each iteration of model training has the same computation logic and the iteration number has little impact on micro-architectural behaviors.  So we sample a partial training set and calculate the FLOPs for efficiency. 
As the  image size of the EWA and ImageNet datasets is 13.14k, and 1280k, respectively, so we sample 500 images and 12800 images from the EWA and ImageNet datasets, respectively.
The throughput is calculate according to the following equation: $Throughput = N \times R \times C$. Here N is the number of images  processed by each training process per second, R is the total number of ranks (the number of training processes), and C is the FLOPs per image.


\begin{table}[ht]
\scriptsize
\centering
\caption{The FLOPs per image.}
\begin{tabular}[ht]{>{\centering}m{0.25\textwidth}|>{\centering}m{0.15\textwidth}>{\centering}m{0.15\textwidth}>{\centering\arraybackslash}m{0.15\textwidth}}
\toprule
\textbf{Dataset}& \textbf{Image Sample Size} & \textbf{Total FLOPs} & \textbf{FLOPs Per Image}\\
\midrule
EWA & 500 & 345.66 TFLOPs & 691 GFLOPs   \\ \midrule
Image Classification & 12800 & 2877.06 TFLOPs & 23 GFLOPs \\ 

\bottomrule
\end{tabular}
\label{table:flop_perimage}
\end{table}%

\begin{table}[ht]
\centering
\scriptsize
\caption {The performance summary of a single node}

\begin{threeparttable}
\begin{tabular}{c|c|c|c|c|c|c}
\toprule
 \textbf{Workloads} &
 \textbf{Models} & \textbf{Precision} & \begin{tabular}{@{}c@{}}
 \textbf{GFLOP}\\
 \textbf{(Per Image)}\\	
 \end{tabular}& \begin{tabular}{@{}c@{}}
 \textbf{Throughput}\\
 \textbf{(Images/s)}\\	
 \end{tabular} 
 &
 \begin{tabular}{@{}c@{}}
 \textbf{Attainable Performance\tnote{1}}\\
 \textbf{(TFLOPS)}\\	
 \end{tabular}  & \begin{tabular}{@{}c@{}}
 \textbf{Achieved Performance Ratio\tnote{2}}\\
 \textbf{(\%)}\\	
 \end{tabular}\\
 \midrule
 Image Classification &
 ResNet-50 V1.5  
 &
 \begin{tabular}{@{}c@{}}
 FP32\\
 Mixed\tnote{3}\\	
 \end{tabular}
 &
 23
 &
 \begin{tabular}{@{}c@{}}
 2624\\
 5734\\
 \end{tabular} 

 &\begin{tabular}{@{}c@{}}
 58\\
 126\\
 \end{tabular} 
   &
 \begin{tabular}{@{}c@{}}
48\\
105\\	
 \end{tabular}\\

\midrule
EWA
&FasterRCNN & FP32 & 691 & 46 & 31 & 26 \\

\bottomrule

\end{tabular}
\begin{tablenotes}
        \footnotesize
        \item[1] The attainable performance refers to the performance obtained in the testing.
        \item[2] The achieved performance ratio refers to the ratio of the attainable performance against the theoretical peak performance (FP32).
        \item[3] Mixed refers to FP32 \& FP16 mixed precision.
            
\end{tablenotes}
\end{threeparttable}
\label{table:summary}
\end{table}

\begin{figure*}[ht]
  
    \includegraphics[width=\linewidth]{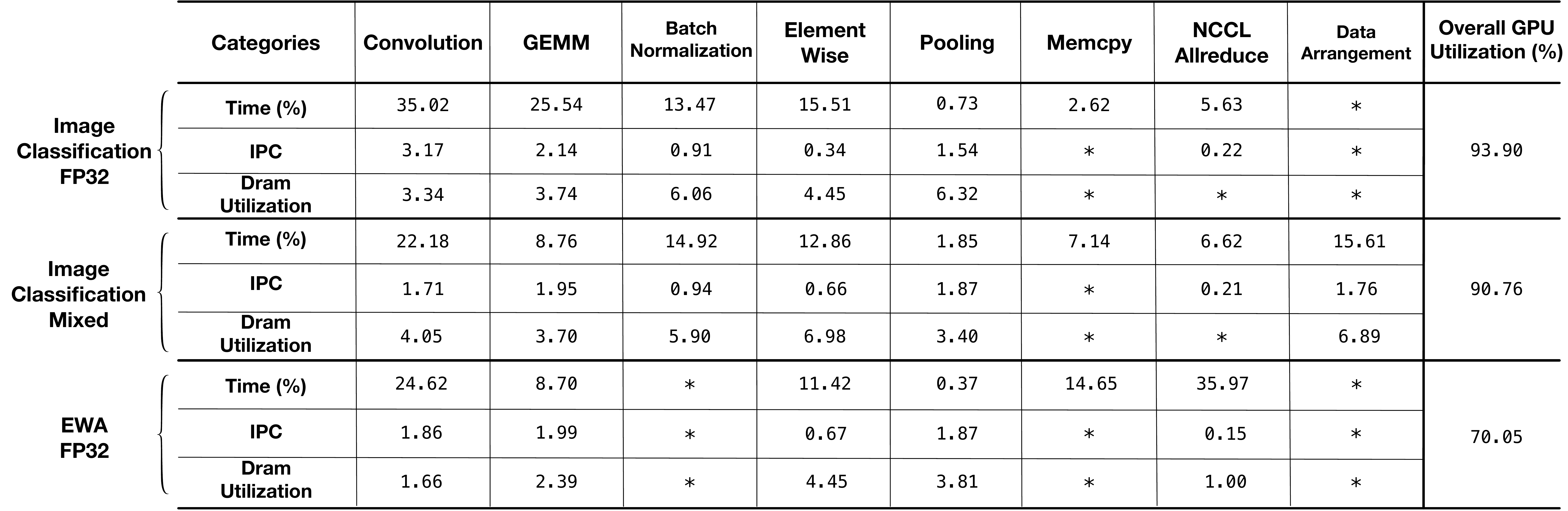}
    \caption{The details of single-node performance analytics of Image Classification and EWA. We classify the kernels invoked on the GPU into eight categories and use three metrics to depict their characteristics: the proportion of time, instruction per cycle (IPC) and dram utilization. The GPU utilization during the overall training session is also recorded. An asterisk (*) is used to indicate the number is negligible, less than 0.001\%.}
    \label{fig:details_single_node}
\end{figure*}

\subsection{Single-node Evaluation} \label{subsec:single-node}

In this subsection, we first report the execution efficiency on a single node, and then perform communication and computation analytics to recover the factors that impact the performance significantly. We use the HPC AI500 V2.0 benchmarks.

\subsubsection{Performance Efficiency} 

Based on the methodology described in Section~\ref{subsec:calculate-flops}, we report the performance efficiency of two benchmarks on a single node: Image Classification and EWA. 
We evaluate both the FP32 precision and mixed precision implementations, which uses the Tensor Core to accelerate the training session.
As the memory footprint required by the mixed precision implementation is nearly a half of that of FP32 precision, we double the batch size in each training step for mixed precision without breaking the benchmarking rule defined in Section~\ref{sec:benchmarking-rules}. Table~\ref{table:summary} shows the performance efficiency of the above two benchmarks. The achieved performance ratio is the ratio of the attainable performance against the theoretical peak performance of the FP32 precision implementation. In our experiments, the  theoretical peak  number is 120 TFLOPS, which is the theoretical peak performance of the  single-precision (FP32) implementation (15 TFLOPS) multiplied by 8-- the number of NVIDIA Tesla V100 SXM2 GPUs. From Table~\ref{table:summary}, we find that the performance efficiency of EWA is extremely low with respect to that of Image Classification. 
We further characterize their computation and communication characteristics to uncover the factors.


\begin{figure}[!ht]
  \centering
  \includegraphics[width=0.8\linewidth]{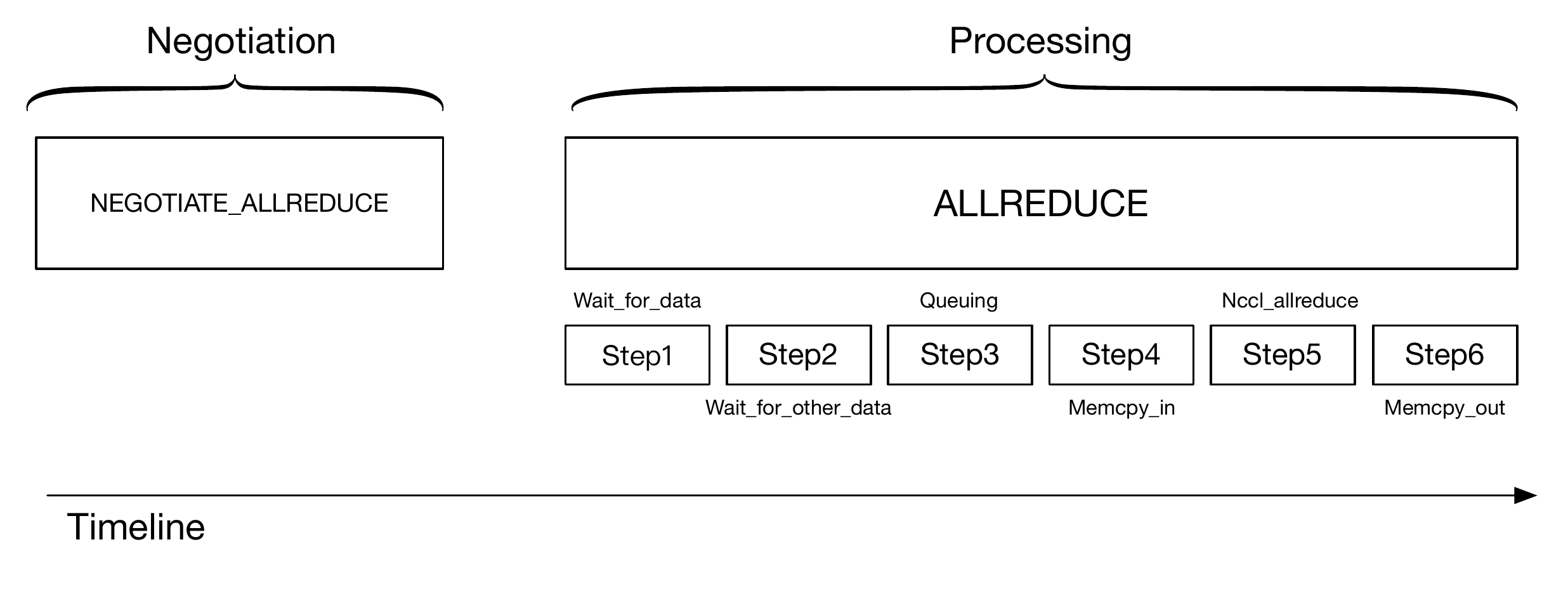}
  \caption{The timeline of Horovod communication. }
  \label{fig:horovod-timeline}
\end{figure}

\subsubsection{Communication and Computation Analytics} 

We first perform communication analytics using a timeline analysis tool ~\cite{horovodtimeline} to record all activities of the Horovod communication, since its synchronous distributed manner may significantly affect the performance. As shown in Fig.~\ref{fig:horovod-timeline},
the communication timeline of Horovod is divided into two phases: ~\emph{negotiation} and ~\emph{processing}. In the ~\emph{negotiation} phase, all training processes send a signal to the first process to ensure their status are ready for the subsequent tensor reduction. In the ~\emph{processing} phase, the tensor reduction is performed. Specifically, the ~\emph{processing} phase is further divided into six steps. Steps 1 (\emph{Wait\_for\_data}) and 2 (\emph{Wait\_for\_other\_data}) are waiting for the data produced by GPU computing, which is the input to all\_reduce operations. Step 3 (\emph{Queuing}) 
happens only when the previous all\_reduce has not finished. Steps 4 (\emph{Memcpy\_in}) copies data into the fusion buffer. Step 5 (\emph{NCCL Allreduce}) is the core part that executes all\_reduce operation across all the training processes.  Steps 6 (\emph{Memcpy\_out}) removes the data out of the fusion buffer.

\begin{table}[ht]

\scriptsize
\centering
\caption{The time breakdown of the Horovod communication.}
\begin{tabular}[ht]{>{\centering}m{0.2\textwidth}|>{\centering}m{0.2\textwidth}|>{\centering}m{0.2\textwidth}|>{\centering\arraybackslash}m{0.2\textwidth}}
\toprule
\textbf{Phases}& \textbf{Steps} &\textbf{EWA} & \textbf{Image Classification} \\
\midrule
Negotiation & Negotiation Allreduce & 54.837 ms & 22.836 ms \\ \midrule
Processing & Wait\_for\_data & 1.746 ms & 85.418 ms \\ \midrule
Processing & Wait\_for\_other\_data & 2.961 ms & 27.036 ms \\ \midrule
Processing & Queuing & 65.863 ms & 0.043 ms \\ \midrule
Processing & Memcpy\_in & 0.108 ms & 1.256 ms \\ \midrule
Processing & NCCL Allreduce & 66.228 ms & 4.153 ms \\ \midrule
Processing & Memcpy\_out & 0.197 ms & 0.993 ms \\ 

\bottomrule
\end{tabular}
\label{table:horovod_time_line}
\end{table}

We profile the average wall clock time of all steps and compare EWA against Image classification. We find the long negotiation phase is one main factor that leads to inefficient communication of EWA. As shown in Table~\ref{table:horovod_time_line}, the average ~\emph{negotiation allreduce} of EWA accounts for 28.5\% of the total duration of Horovod communication, 2.5 times than that of Image classification. The root cause is the side effect of  the centralized schedule strategy of the Horovod negotiation. As mentioned before, the first process during the negotiation acts as a centralized scheduler to avoid deadlock by reordering all the all\_reduce operations across processes. It receives the message from all processes and sends back the correct tensor list that should be reduced. EWA needs to execute all\_reduce operation more than one hundred times and has about 41 millions of gradients in total to be reduced during each training step, and thus spends too much time on the first process. Another factor is the sub-optimal overlap between computation and communication. According to Table~\ref{table:horovod_time_line}, 
we find the total duration of ~\emph{wait\_for\_data} and ~\emph{wait\_for\_other\_data} in EWA and Image Classification is 4.6, 112.4 ms, respectively; in  the duration of ~\emph{NCCL 
Allreduce} it is 66.2 ms and  4.15 ms in  EWA and Image Classification, respectively. These numbers indicate EWA has a worse overlap between computation and communication than that of Image classification. Besides, ~\emph{queuing} is up to about 65.8 ms, showing the  ~\emph{NCCL Allreduce} operation has to wait for a longer duration. 
Accordingly, the duration of ~\emph{queuing}  and  ~\emph{wait\_for\_data} of Image Classification is 0.043 and 85.4 ms, respectively, indicating Image Classification has better overlap between communication and computation than that of EWA.

In addition to the communication analytics, we also conduct computation analytics through a thorough profiling of GPU activities using NVProf~\cite{nvprof}.  Fig.~\ref{fig:details_single_node} shows the results. There are thousands invocations of CUDA kernel  during each training step. For simplicity, we classify all the kernel functions into eight categories. Each category represents a kind of operation, namely convolution, GEMM, batch normalization, element wise, pooling, memcpy, NCCL Allreduce, and transformation. For EWA, we find NCCL Allreduce (35.97\%) and memcpy operations occupy 50.62\% in total, leading to poor performance. For ImageNet Classification, the most time-consuming kernel is convolution, namely 35.02\% and 22.18\% in the FP32 and mixed precision implementations, respectively. 

We also notice the overhead of data arrangement occupies 15.61\% in the mixed precision implementation, while less than 0.0001\% in the FP32 implementation. The huge overhead in the mixed precision implementation is incurred by converting different data layouts between the TensorFlow and CUDA kernels. The data layout of the TensorFlow kernels is represented in a quadruple tuple (batch size, channels, height of data sample, width of data sample),  abbreviated as NCHW.  
While, the data layout of the CUDA kernels is represented in  a quadruple tuple (batch size, height of data sample, width of data sample, channels), abbreviated as NHWC. That inconsistency incurs a huge overhead.  
It explains why the speedup of mixed precision version of Image Classification is only 2.16x. It is much smaller than the results published by Nvidia~\cite{nvidiav100}, which claims that the mixed precision training can bring up to 8x speed up on the Tesla V100 GPU. 

\captionsetup[figure]{font=scriptsize}
\begin{figure*}[ht]
\begin{subfigure}{0.33\textwidth}
  \centering
  \includegraphics[width=.9\linewidth]{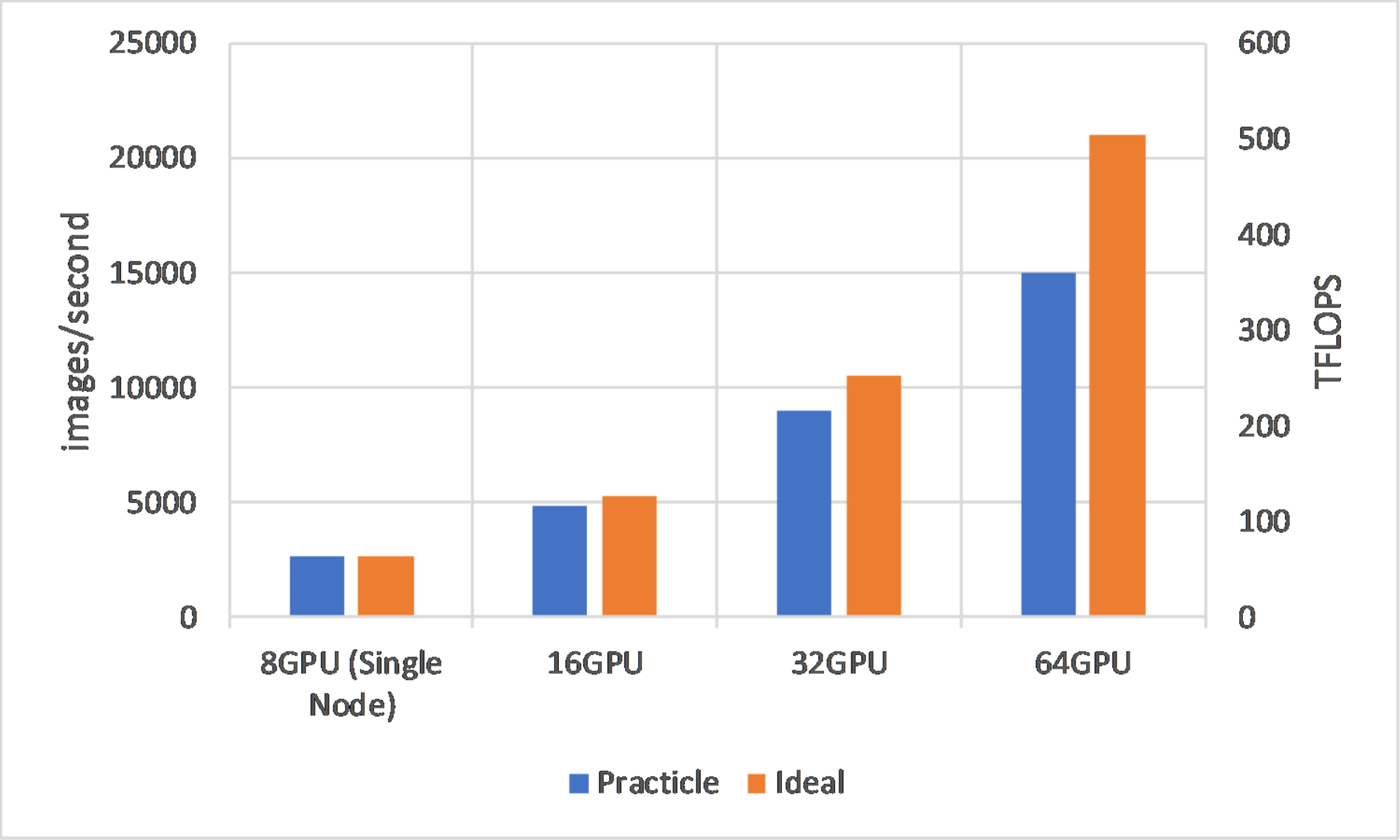}
  \caption{Image Classification (FP32) }
  \label{fig:imagenet-fp32-scaling}
\end{subfigure}
\begin{subfigure}{0.33\textwidth}
  \centering
  \includegraphics[width=.9\linewidth]{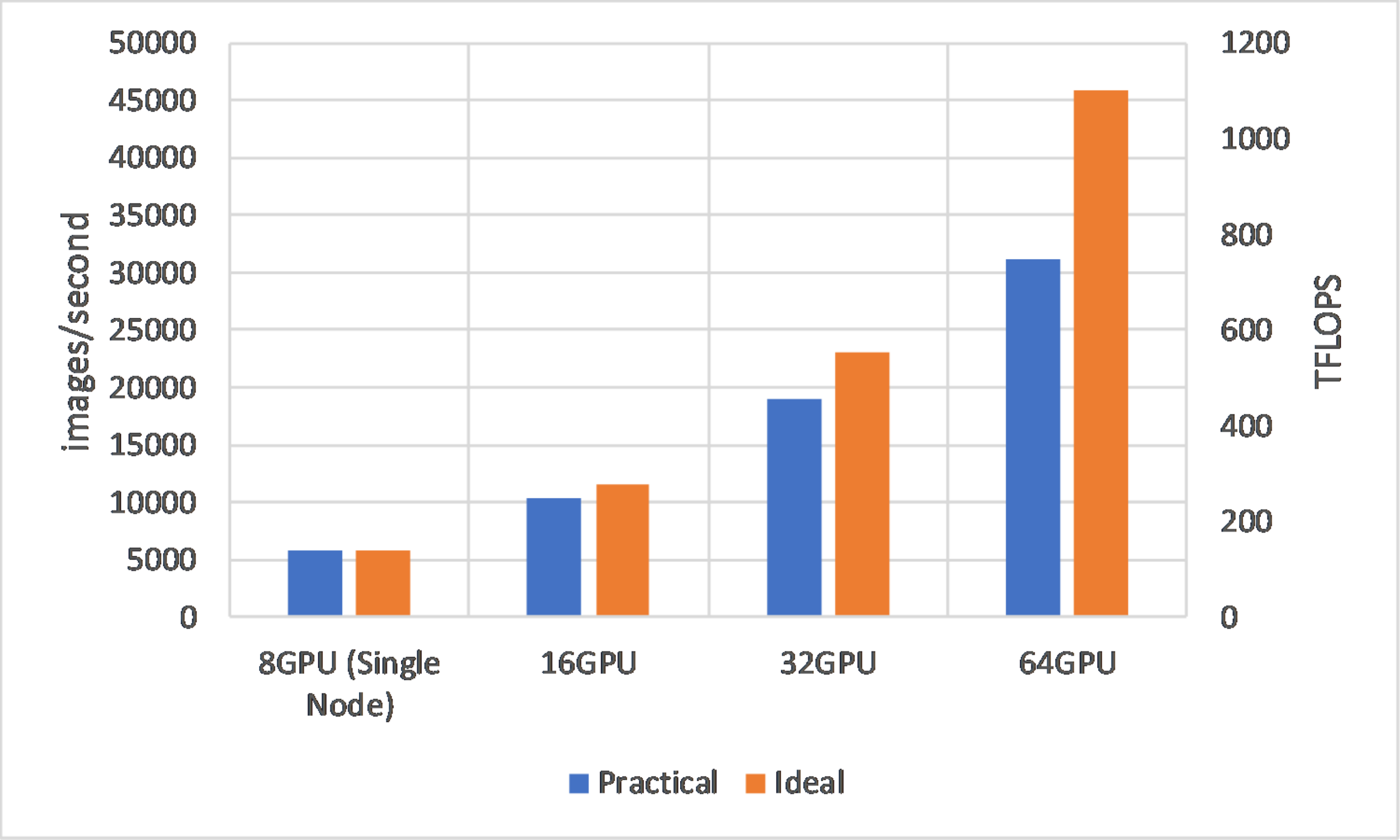}
  \caption{Image Classification (Mixed)}
  \label{fig:imagenet-mixed-scaling}
\end{subfigure}
\begin{subfigure}{0.33\textwidth}
  \centering
  \includegraphics[width=.9\linewidth]{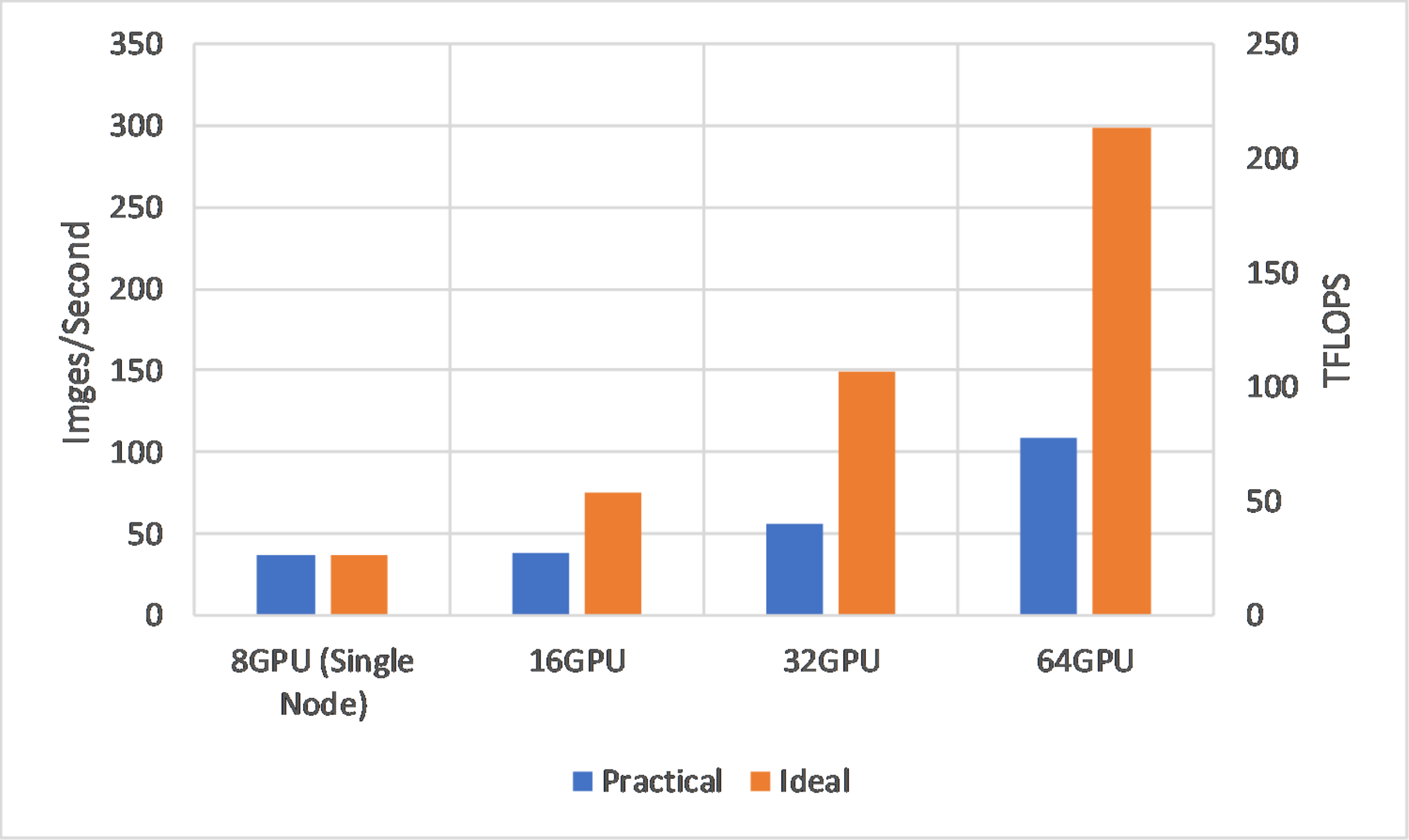}
  \caption{EWA (FP32)}
  \label{fig:EWA-fp32-scaling}
\end{subfigure}
\medskip
\begin{subfigure}{0.33\textwidth}
  \centering
  \includegraphics[width=.9\linewidth]{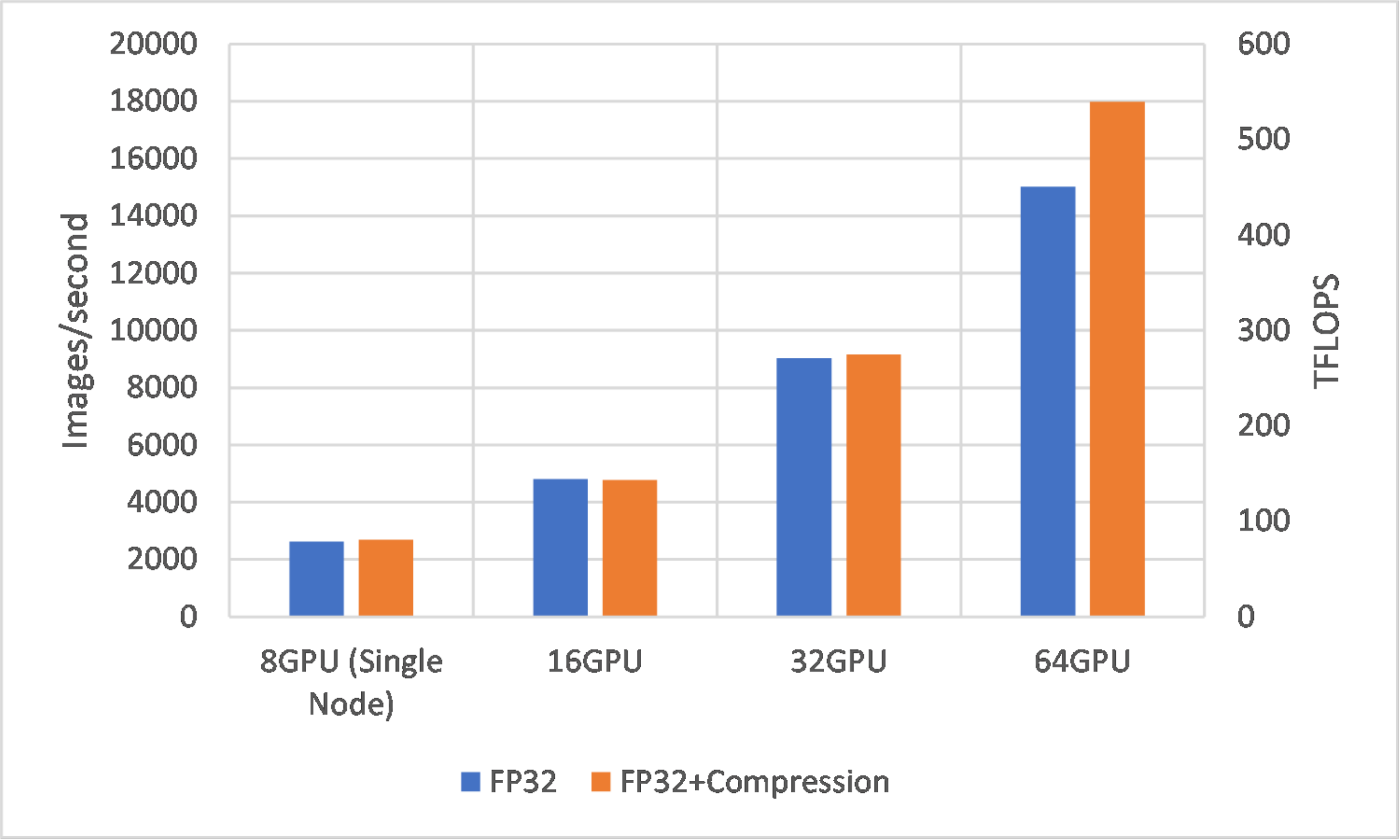}
  \caption{Image Classification (FP32+Compression) }
  \label{fig:imagenet-fp32-compression-scaling}
\end{subfigure}
\begin{subfigure}{0.33\textwidth}
  \centering
  \includegraphics[width=.9\linewidth]{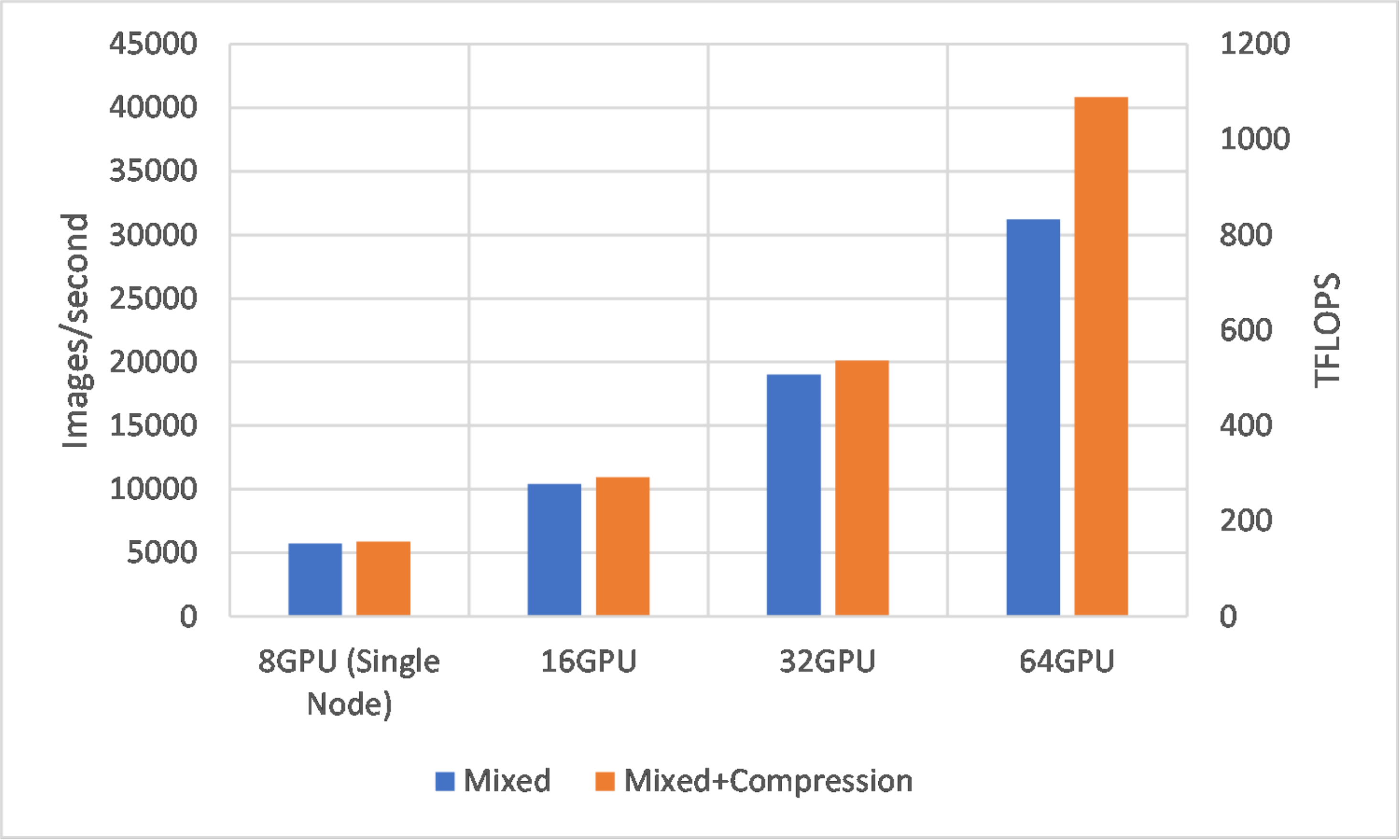}
  \caption{Image Classification (Mixed+Compression)}
  \label{fig:imagenet-mixed-compression-scaling}
\end{subfigure}
\begin{subfigure}{0.33\textwidth}
  \centering
  \includegraphics[width=.9\linewidth]{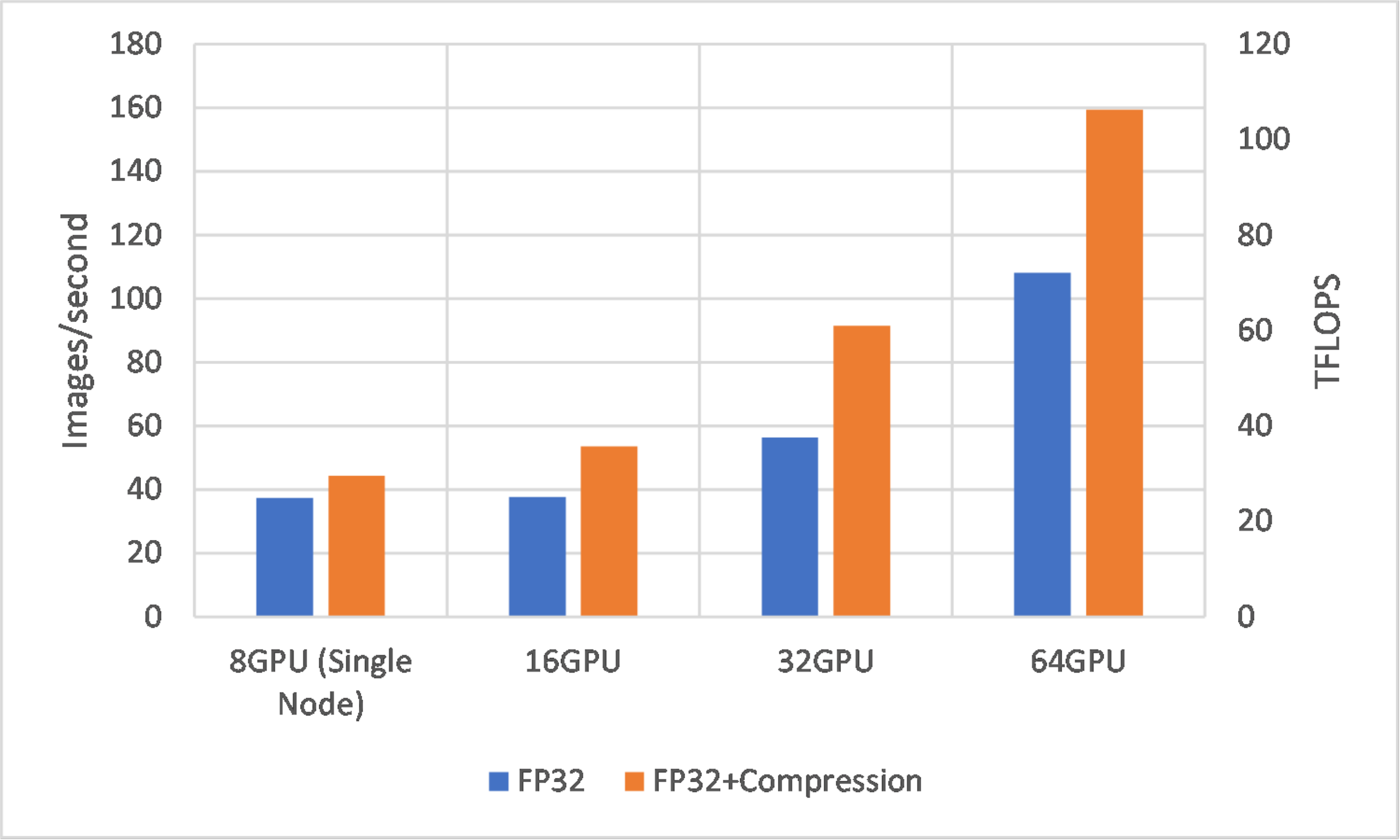}
  \caption{ EWA (FP32+Compression)}
  \label{fig:EWA-fp32-compression-scaling}
\end{subfigure}
\caption{The scaling experiments of EWA and Image 
Classification. 
}
\label{fig:scaling}
\end{figure*}

\captionsetup[figure]{font=normalsize}

\subsection{Multiple-node Evaluation}\label{subsec:scaling-expriments} 


We perform several scaling experiments on the distributed system, described in Section~\ref{subsec:ex-config}. Both EWA and Image Classification experiments are scaled out from 8 GPUs to 64 GPUs. We take the 8-GPU experiments (single node) as a baseline. Our communication topology is the double binary tree~\cite{NCCL}, which is implemented by NCCL 2.4. We report the performance numbers of these experiments and perform  further analysis using the HPC AI roofline models proposed in Section~\ref{subsec:roofline-model}. The scaling results are shown in Fig.~\ref{fig:scaling}.

\subsubsection{Image Classification} 
 For the FP32 precision implementation of Image Classification, the parallel efficiency is 0.91, 0.85 and 0.71 on 16, 32 and 64 GPUs, respectively. For the mixed implementation, the parallel efficiency is slight lower: 0.89, 0.82 and 0.67, respectively. There is a notable loss of parallel efficiency when the system scale is 64 GPUs. 
 
 We also notice that the communication compression does not bring any performance improvement when the system scale is 32 GPUs or less. However, when the scale is 64 GPUs, it contributes a lot. For the FP32 version, the performance improves from 345 to 414 TFLOPS. For the mixed version, the performance  improves from 718 to 939 TFLOPS.
 According to our HPC AI Roofline model shown in Fig.~\ref{fig:DisNodeRoof}, we find that there is a performance bound shift when the system scale changes from 32 to 64 GPUs. Specifically, when the system scale is less or equal to 32 GPUs, Image Classification's communication ceiling is dominated by NVLink's bandwidth, and it is computation-bound. Hence, communication compression cannot improve the performance. However, When the system increases to 64 GPUs, the communication ceiling is dominated by Ethernet's bandwidth, so it turns into communication-bound. This is why communication compression works. The highest performance of Image Classification that we achieve is 939 TFLOPS through both mixed precision optimization and communication compression, as shown in Fig.~\ref{fig:imagenet-mixed-compression-scaling}.    
 
\subsubsection{EWA}
For the FP32 precision implementation of EWA, the parallel efficiency is 0.50, 0.37, and 0.36 at the system scale of 16, 32, and 64 GPUs, respectively. According to the Roofline model shown in Fig.~\ref{fig:DisNodeRoof}, the bottleneck is always communication bandwidth. Therefore, communication compression achieves good results. When communication compression is used, the performance gain persists when the scale increase from 8, to 16, 32, and 64 GPUs, and the speedup is 1.2, 1.4 ,1.6 and 1.5, respectively. The highest performance of EWA achieved through communication compression is 109 TFLOPS. 

\begin{figure}[ht]
    \centering
    \includegraphics[width=.6\linewidth]{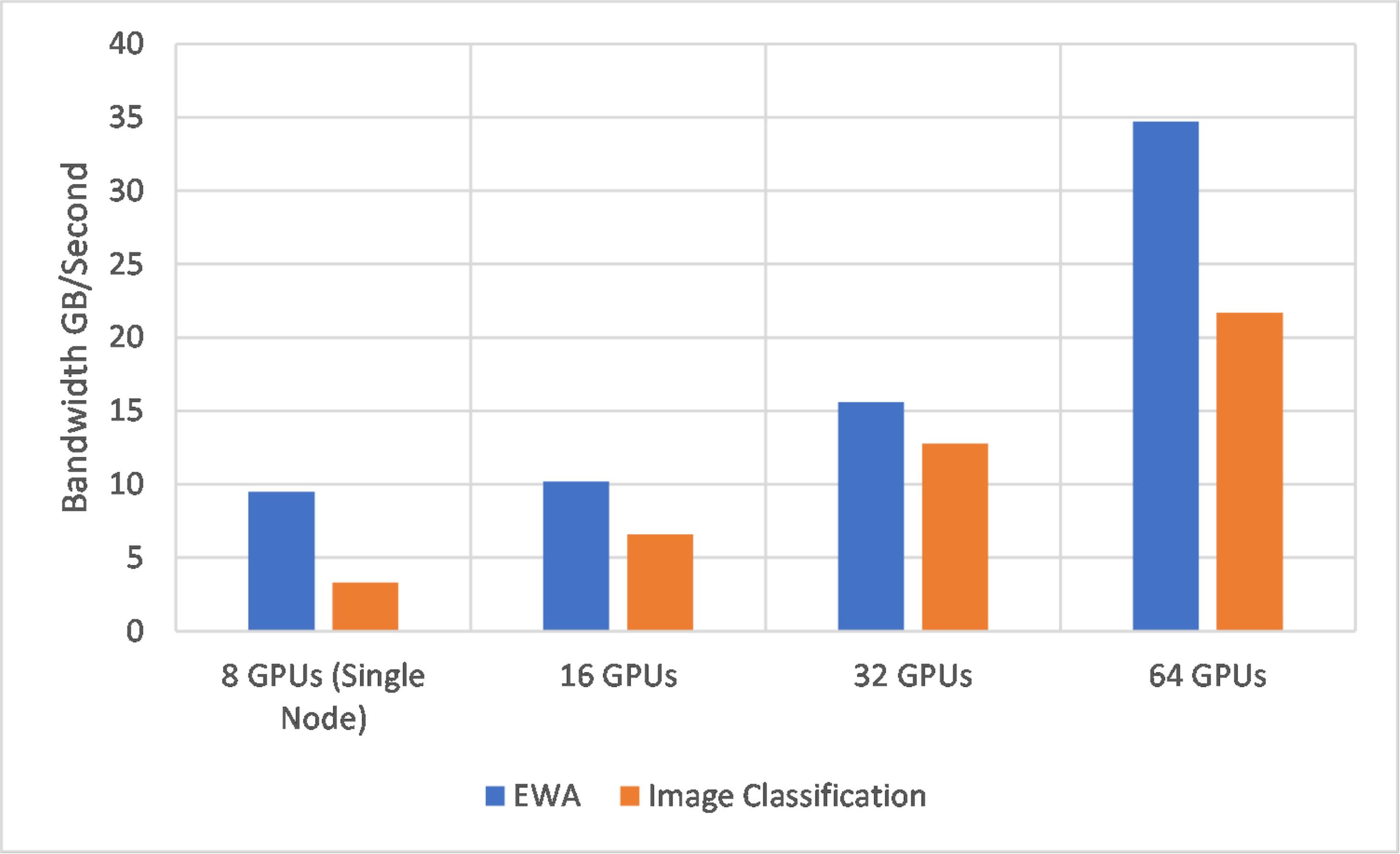}
    \caption{The distinctive communication bandwidth consumption of the FP32 implementations of EWA and Image Classification. }
    \label{fig:bandwidth-ewa-imagenet}
\end{figure}

\subsubsection{Why EWA and Image Classification Have Different Parallel Efficiencies?}

For EWA and Image Classification, we found their different parallel efficiencies are due to distinct communication bandwidth consumption. As shown in Fig.~\ref{fig:bandwidth-ewa-imagenet}, we measure the communication bandwidth consumption of the FP32 precision implementations of EWA and Image Classification. EWA consumes much higher communication bandwidth than that of Image Classification. In the contrast, the performance of Image Classification largely depend on the computation efficiency especially when the scale is less than and equal to 32 GPUs. In conclusion, 10 Gb/s Ethernet can not satisfy the communication requirement of EWA, and hence results in poor parallel efficiency.

\subsubsection{The VFLOPS Ranking of HPC AI Systems Using Image Classification}

\begin{figure}[ht]
  \centering
  \includegraphics[width=0.6\linewidth]{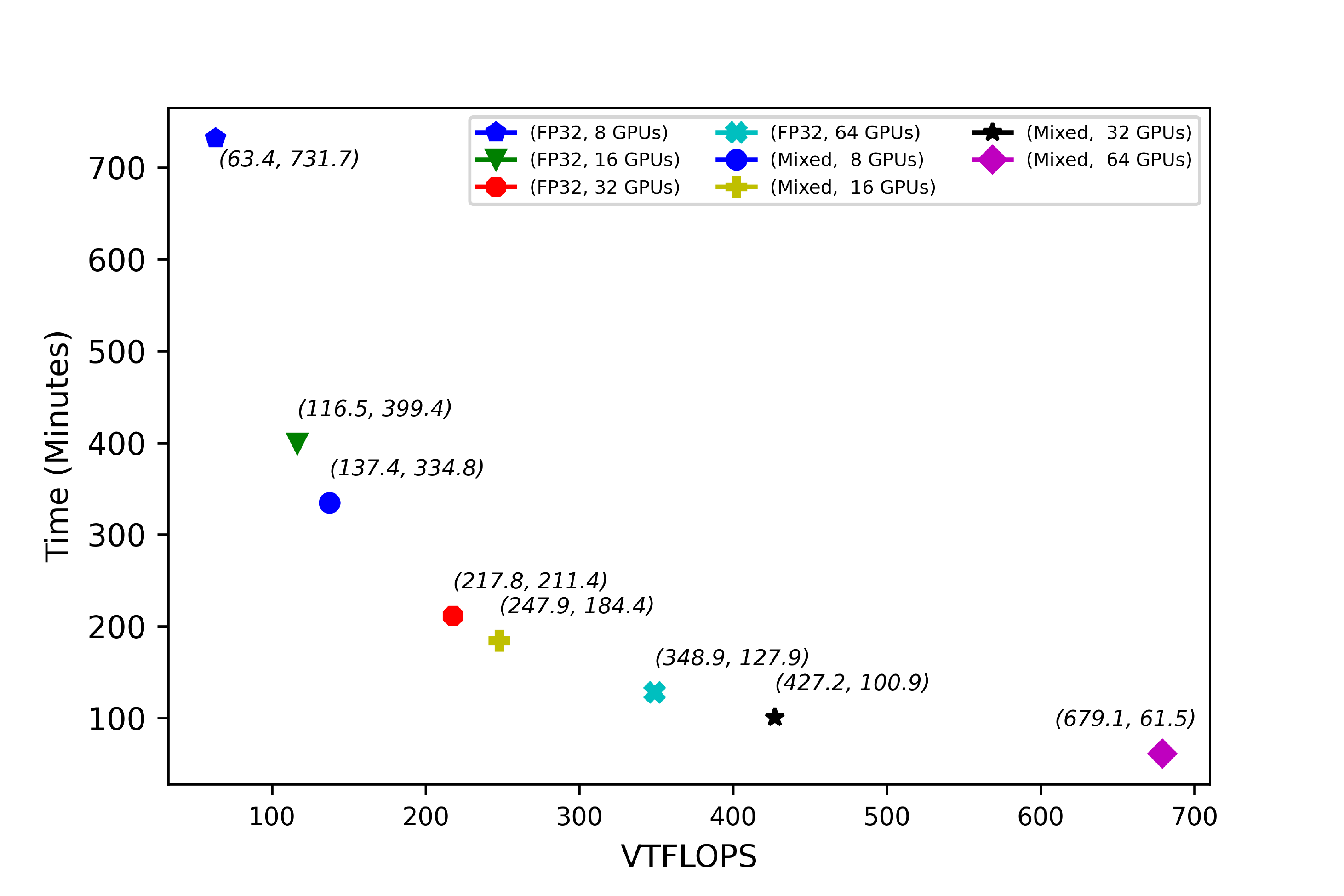}
  \caption{The VFLOPS Rankings of HPC AI Systems Using Image Classification. }
  \label{VFLOPS_ranking}
\end{figure}

The metric of  VFLOPS emphasizes both the performance and quality. 
Fig.~\ref{VFLOPS_ranking} shows the rankings of different scale HPC AI systems with mixed-precision or FP32 implementations. The highest performance is 642 TVFLOPS, achieving through the mixed optimization at the scale of 64 GPUs. Meanwhile, another auxiliary metric--time-to-quality is also reported. Generally, our metric is simple and visual. 


\subsection{The Case Study of Using HPC-AI Roofline Models} \label{subsec:roofline-model}

This section presents a case study on how to use our proposed HPC AI Roofline models to identify the bottleneck and guide optimizations. 

\subsubsection{Bottleneck Identification}

We use the proposed roofline models to the 16-GPU HPC AI system. The theoretical peak  number is calculated according to the hardware configurations shown in Table~\ref{table:hwconfigeration}. We use the roofline model to identify potential bottlenecks of EWA and Image Classification.

\begin{figure}[ht]
  \centering
  \includegraphics[width=0.6\linewidth]{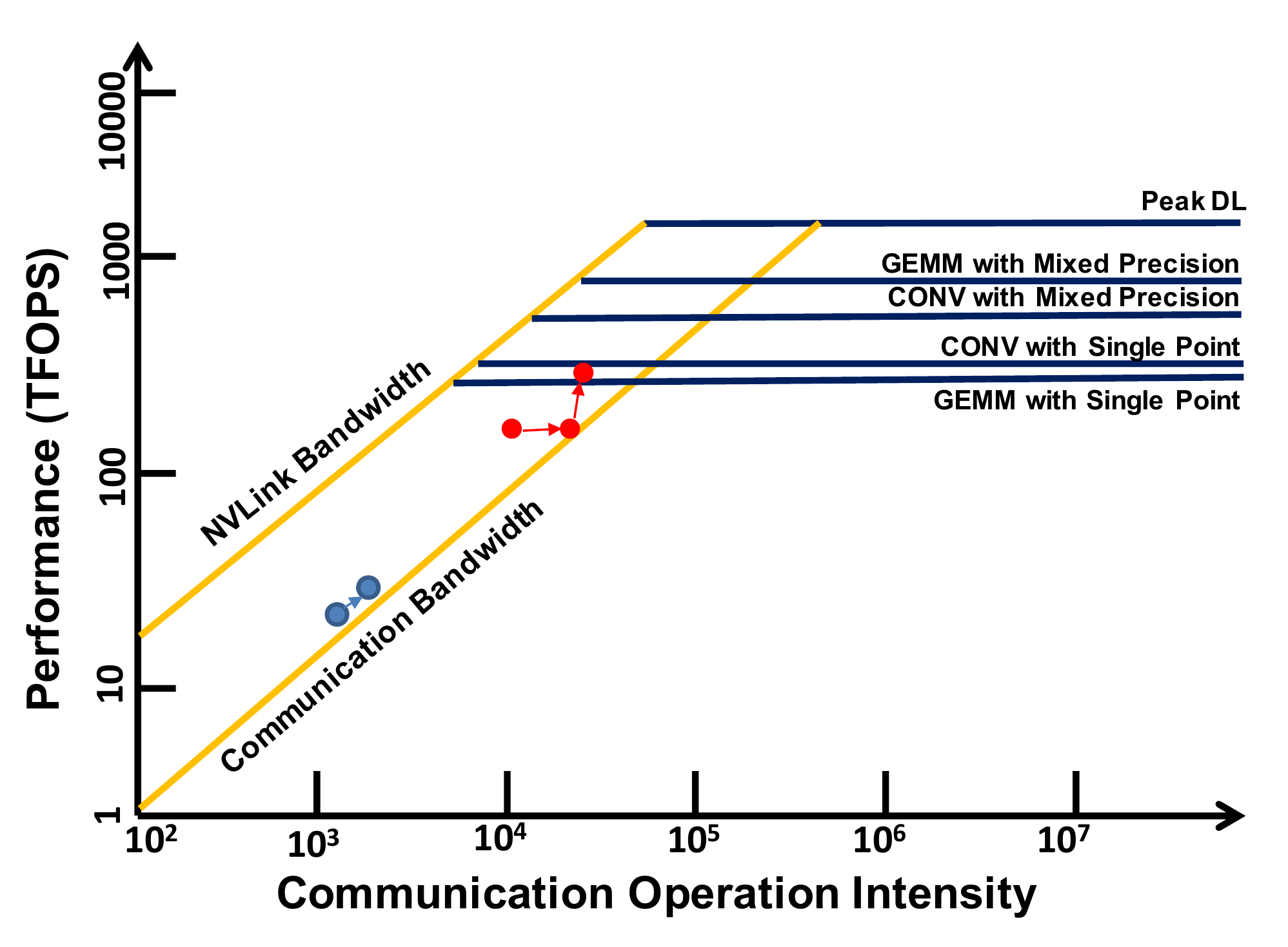}
  \caption{The roofline model at the system scale of 16 GPUs. The blue point represents EWA, and the red point represents Image Classification. }
  \label{fig:2nodes-usecase-roofline}
\end{figure}

From Fig.~\ref{fig:2nodes-usecase-roofline}, we have the following observations. EWA is bounded by the communication bandwidth as it falls in the slanted part of the roof, while Image Classification is bounded by the computation as it falls in the flat part. 


\subsubsection{Optimizations}
We adopt two optimization strategies: communication compression and mixed precision optimization.
\paragraph{Communication compression.} In order to optimize the communication, we perform communication compression, which encodes and compresses the tensor precision into FP16 for communication and then decodes into FP32 for computation. This optimization halves the amount of communication for each training step, which is equivalent to doubling the communication bandwidth. As the amount of computation remains the same, the COI of EWA and Image Classification also doubles. As shown in Fig.~\ref{fig:2nodes-usecase-roofline}, our results show that the performance of EWA increases from 25.99 to 36.97 TFLOPS after communication compression.
On the other hand, the performance of Image Classification is not improved because it is computation bound. Its COI indeed increases.

\paragraph{Mixed precision training.} In order to improve the performance of Image Classification, we adopt the mixed precision optimization, which makes use of Tensor Core to perform arithmetic calculation in an FP16 format, achieving higher amount of computation operations per second. As shown in  Fig.~\ref{fig:2nodes-usecase-roofline}, the rightest red point represents using the mixed precision training.  It brings about 2.16x speedup. Moreover, the COI is also improved. This is because that the mixed precision training requires lower memory footprint, so we double the batchsize, and the larger batchsize leads to higher COI (higher amount of computation per step). In the near future, we will try mixed precision optimization for EWA, too.

\section{Related Work}\label{related_work}

We summarize the related work in a chronological order (according to the publication dates of the referred papers or publicly available technique reports) from the perspectives of HPC benchmarking, AI benchmarking, and HPC AI benchmarking.



\subsection{HPC Benchmarking}

HPL (1994)~\cite{dongarra2003linpack} is the famous HPC benchmark for the Top500~\cite{dongarra1997top500} ranking. HPL is short for High Performance Linpack,  which is designed to solve dense linear equations. For the TOP500 ranking, users are allowed to optimize MPI~\cite{openmpi} and the BLAS~\cite{blas} library to achieve the best performance.
Since solving the  Linpack problem  is very regular,  the HPC system can achieve very high performance.  So, the performance of HPL can be described as the upper bound performance of the target HPC system. 
HPL is open source and publicly available from~\url{https://www.netlib.org/benchmark/hpl/} .

NPB (1994)~\cite{npb} is the NAS Parallel Benchmark suite, whose workloads are derived from the computational fluid dynamics (CFD) applications. CFD is a typical traditional HPC application. Based on the  pencil-and-paper specification, NPB 1.0 consists of five kernels and three pseudo-applications, and the lastest NPB 3.4.1 includes 12 workloads. NPB is open source and  publicly available from~\url{https://www.nas.nasa.gov/publications/npb.html} .

HPCC (2005)~\cite{humphrey2009evaluating} is an HPC Challenge benchmark suite, which includes seven different workloads. HPCC covers the spectrum of spatial locality and  temporal locality of the HPC workloads. So, the HPCC benchmarks are designed for measuring a range of memory access patterns of the HPC system. HPCC is open source and  publicly available from~\url{https://icl.utk.edu/hpcc/} .

Graph500 (2010)~\cite{ueno2012highly} is designed for the data-intensive supercomputer applications. The workloads of Graph500 are the search and  shortest-path programs of the  weighted undirected graph. The Graph500 workloads exhibit very low spatial and temporal locality. Its metric is not the FLOPS but the TEPS (traversed edges per second). 
Graph500 is open source and publicly available from~\url{https://graph500.org/} .

HPCG (2013)~\cite{dongarra2016high} is another benchmark for the Top500 ranking. HPCG means High Performance Conjugate Gradients (HPCG). Computational and data access patterns of HPCG are more close to the real HPC applications. As a kernel workload  extracted from the traditional HPC workloads, the HPCG benchmark is intended as a complement to the High Performance LINPACK (HPL) benchmark, and the FLOPS of HPCG is far lower than that of HPL on the same platform. 
HPCG is open source and publicly available from~\url{https://github.com/hpcg-benchmark/hpcg} .

\subsection{AI Benchmarking}

BenchNN (2012)~\cite{chen2012benchnn} uses neural networks algorithms to re-implement the well-known PARSEC benchmark~\cite{bienia2008parsec}. Their main propose is to illustrate the potential application scope of neural networks algorithms. The models adopted in BenchNN are simple shallow neural networks such as multi-layer perceptron, and  thus they cannot reflect the state of the art. BenchNN is not open source so far.

DeepBench (2016)~\cite{deepbench} is a micro benchmark suite that aims to benchmark basic operations in deep neural networks such as convolution and dense matrices multiply. The methodology of DeepBench is to reflect the characteristics of these operations by using different input sizes.
Since only operator level is concerned, DeepBench cannot provide full-model level evaluation. DeepBench is open source and publicly available from~\url{https://github.com/baidu-research/DeepBench} .


Both Fathom (2016)~\cite{adolf2016fathom} and TBD (2018)~\cite{zhu2018benchmarking} consists of representative AI workloads, covering a broad range of application domains. Their evaluation only focus on throughput while ignoring model quality. Fathom is open source and publicly available from \url{https://github.com/rdadolf/fathom} . TBD is open source and publicly available from \url{https://github.com/tbd-ai/tbd-suite} .

DawnBench (2017) ~\cite{coleman2017dawnbench} aims to end-to-end deep learning benchmarking as it firstly proposes time-to-accuracy as the main metric, which requires to train a model to the state-of-the-art accuracy. It has two workloads including image classification and question answering. The limitation of DawnBench is ignoring the equivalent benchmarking rules. 
DawnBench is open source and publicly available from \url{https://github.com/stanford-futuredata/dawn-bench-entries} .

The BenchCouncil AI benchmark suites (2018) present a series of AI benchmarking work, including AIBench~\cite{gao2018aibench,gao2019aibench,gao2020aibench,tang2020aibench} for datacenter AI benchmarking, AIoTBench~\cite{luo2018aiot} for mobile and embedded device intelligence benchmarking, Edge AIBench~\cite{hao2018edge} for edge computing benchmarking, and the previous version of HPC AI500~\cite{jiang2018hpc}. The BenchCouncil AI benchmarks are by far the most comprehensive AI benchmark suites covering datacenter, IoT, edge, and HPC. For example, AIBench adopts a scenario-distilling benchmarking methodology for the first time, which considers scenario benchmarks, component and micro benchmarks as three indispensable parts of a benchmark suite. This methodology bridges a huge gap from real-world application deployments to simulator-based architecture research, and balances the subtly different requirements of earlier-stage benchmarking (portability and affordability for new architectures) and later-stage benchmarking (representativeness and comprehensiveness)~\cite{gao2020aibench}.
The BenchCouncil AI benchmark suites are open source and publicly available from~\url{http://www.benchcouncil.org/benchmarks.html}.

BenchIP (2018)~\cite{tao2018b} focuses on benchmarking intelligent processors. It contains two sets of benchmarks:  micro-benchmarks and macro-benchmarks. Micro-benchmarks consists of single-layer networks that are used to system optimizations. Macro-benchmarks consists of various neural networks that are used to offer realistic benchmarking. BenchIP also ignores the equivalent benchmarking rules. In addition, it only focuses on throughput. BenchIP is not open source so far.


MLPerf (2019)~\cite{mattson2019mlperf} includes seven benchmarks for training and five benchmarks for inference. The MLPerf training benchmark proposes a series of benchmarking rules to eliminate the side effect of the stochastic nature of AI. Nevertheless, The MLPerf rules can not be used to assure the equivalency, repeatability, and replicability of HPC AI benchmarking. 
It lacks the specific parallelism and communication rules. MLPerf is open source and publicly available from~\url{https://github.com/mlperf} .

\subsection{HPC AI Benchmarking}

HPC AI500 (V 1.0) (2018)~\cite{jiang2018hpc} is the first HPC AI benchmarks based on the real-world scientific dataset, covering three representative HPC AI applications, namely high energy physics, cosmology, and extreme weather analytics. The HPC AI500 (V 1.0) is open source and publicly available from~\url{http://www.benchcouncil.org/benchhub/hpc-ai500-benchmark} . 

The HPL-AI benchmark (2019)~\cite{hplai} is designed for 32-bit and even lower floating-point precision AI computing.  Using the solver formulation of the decades-old HPL framework of benchmarking,  HPL-AI strives to unite traditional HPC and state-of-art AI.  HPL-AI algorithm is a combination of low-precision (state-of-art AI precision)  LU factorization and iterative refinement performed afterwards to bring the solution back to 64-bit accuracy (traditional HPC precision). However, the LU factorization operation is irrelevant to most of AI workloads. As a micro-benchmark, HPL-AI benchmark is more suitable for evaluating the upper bound performance of the HPC AI system. The HPL-AI benchmark is open source and publicly available from~\url{https://icl.bitbucket.io/hpl-ai/} .

Deep500 (2019)~\cite{ben2019modular} is a reproducible customized benchmarking infrastructure for high-performance deep learning. It has four levels of abstraction to provide a full-stack evaluation. However, its reference implementation uses commercial open source data sets and simple deep learning models, hence cannot reflect real-world HPC AI workloads. Moreover, it fails to propose rules to assure the equivalency, repeatability, and replicability of HPC AI benchmarking.  Deep500 is open source and publicly available from~\url{https://github.com/deep500/deep500} . 

AAH (2020)~\cite{aah} uses AutoML~\cite{jin2019auto} to benchmark HPC AI systems. AutoML is  highly compute-intensive and extensible, which fits the requirement of benchmarking  HPC systems. However, as a complicated AI workload, AutoML involves many hyper-parameters, which usually makes it hard to evaluate~\cite{yang2019evaluation}. Moreover, the variance of its essential workload--Neural Architecture Search is also high as 6.15\%, according to the evaluation in~\cite{tang2020aibench}.

\subsection{ImageNet/ResNet50 Training Optimization}

Some specific AI workloads also play an important role in evaluating the HPC AI system. ImageNet/Resnet50 is a well-known showcase for optimizing HPC AI systems, motivating a series of researches on learning rate scheduling algorithms and efficient communication strategies~\cite{you2017imagenet, goyal2017accurate, akiba2017extremely, jia2018highly, 10.1145/3225058.3225069, yamazaki2019yet,cho2017powerai}. 

The researchers of Facebook (2017)~\cite{goyal2017accurate} formally propose linear scaling rule and warmup schema for the first time and summarized several pitfalls in large scale deep learning. They finish the training in 60 minutes with a top1 accuracy of 76.3\%. 

The researchers of Berkeley (2017)~\cite{you2017imagenet} firstly propose LARS (Layer-wise Adaptive Rate Scaling), which is a novel learning rate policy. By utilizing this policy, they successfully scale the batchsize of ResNet50 to 32K and reduce the training time to 20 minutes. 

Preferred Networks (2017)~\cite{akiba2017extremely}, IBM (2017)~\cite{cho2017powerai}, Tencent (2018)~\cite{jia2018highly}, Sony (2018)~\cite{tanakaimagenet}, Google (2018)~\cite{ying2018image}, and Fujitsu (2019)~\cite{yamazaki2019yet} all focus on high efficient communication strategies (scale to larger HPC systems) and other system-level optimizations (e.g. mixed precision training). Their learning rate policy or other algorithm-level optimizations  follow the work from Facebook~\cite{goyal2017accurate} and Berkeley~\cite{you2017imagenet}. These work have reduced the training times from hours to  minutes. So far, the fastest training time is 74 seconds, which is from Fujitsu (2019)~\cite{yamazaki2019yet}.

\section{Conclusion}

This paper proposes a comprehensive HPC AI benchmarking methodology that achieves the goal of being equivalent, relevant, representative, affordable, and repeatable. Following this methodology, we present open-source benchmarks, and Roofline performance model to benchmarking and optimizing the systems. We propose two innovative metrics:  Valid FLOPS, and valid FLOPS per watt to rank the performance and energy-efficiency of HPC AI systems. 

The evaluations show our methodology, benchmarks, performance models, and metrics can measure, optimize, and rank  the HPC AI systems in a scalable, simple, and affordable way. 
The specification, source code, and benchmarking data are publicly available from \url{http://www.benchcouncil.org/benchhub/hpc-ai500-benchmark} .

\section{Acknowledgments}
We thank the PengCheng Laboratory for hardware support. We also thank Shaomeng Cao, Xuhui Shao, Yongheng Liu, Changsong Liu, and Jingfei Qiu for technical support in using those systems.

\end{document}